\documentclass[10pt,twocolumn,letterpaper]{article}

\usepackage[pagenumbers]{cvpr}

\usepackage{multirow}
\usepackage{tabularx}
\usepackage{colortbl}
\usepackage{mathcomp}
\usepackage[accsupp]{axessibility}

\newcolumntype{C}{>{\centering\arraybackslash}X}
\newcolumntype{L}{>{\raggedright\arraybackslash}X}
\newcolumntype{R}{>{\raggedleft\arraybackslash}X}

\newcommand{\first}[1]{\multicolumn{1}{>{\columncolor[rgb]{1.0,0.8,0.8}}c}{#1}}
\newcommand{\second}[1]{\multicolumn{1}{>{\columncolor[rgb]{1.0,0.9,0.7}}c}{#1}}
\newcommand{\third}[1]{\multicolumn{1}{>{\columncolor[rgb]{1.0,1.0,0.6}}c}{#1}}
\newcommand{\gray}[1]{\textcolor[rgb]{0.5,0.5,0.5}{#1}}
\newcommand\textstrong[1]{\textcolor{Magenta}{\textbf{\textit{#1}}}}

\definecolor{cvprblue}{rgb}{0.21,0.49,0.74}
\usepackage[pagebackref,breaklinks,colorlinks,allcolors=cvprblue]{hyperref}

\renewcommand{\baselinestretch}{0.96}

\def\paperID{10852}
\def\confName{CVPR}
\def\confYear{2025}

\title{Structure from Collision}

\author{Takuhiro Kaneko\\
  NTT Corporation}

\begin{document}
\maketitle

\begin{abstract}
  Recent advancements in neural 3D representations, such as neural radiance fields (NeRF) and 3D Gaussian splatting (3DGS), have enabled the accurate estimation of 3D structures from multiview images. However, this capability is limited to estimating the visible external structure, and identifying the invisible internal structure hidden behind the surface is difficult. To overcome this limitation, we address a new task called Structure from Collision (SfC), which aims to estimate the structure (including the invisible internal structure) of an object from appearance changes during collision. To solve this problem, we propose a novel model called SfC-NeRF that optimizes the invisible internal structure of an object through a video sequence under physical, appearance (i.e., visible external structure)-preserving, and keyframe constraints. In particular, to avoid falling into undesirable local optima owing to its ill-posed nature, we propose volume annealing; that is, searching for global optima by repeatedly reducing and expanding the volume. Extensive experiments on 115 objects involving diverse structures (i.e., various cavity shapes, locations, and sizes) and material properties revealed the properties of SfC and demonstrated the effectiveness of the proposed SfC-NeRF.\footnote{\label{foot:samples}The project page is available at \url{https://www.kecl.ntt.co.jp/people/kaneko.takuhiro/projects/sfc/}.}
\end{abstract}

\begin{figure}[t]
  \centering
  \includegraphics[width=\linewidth]{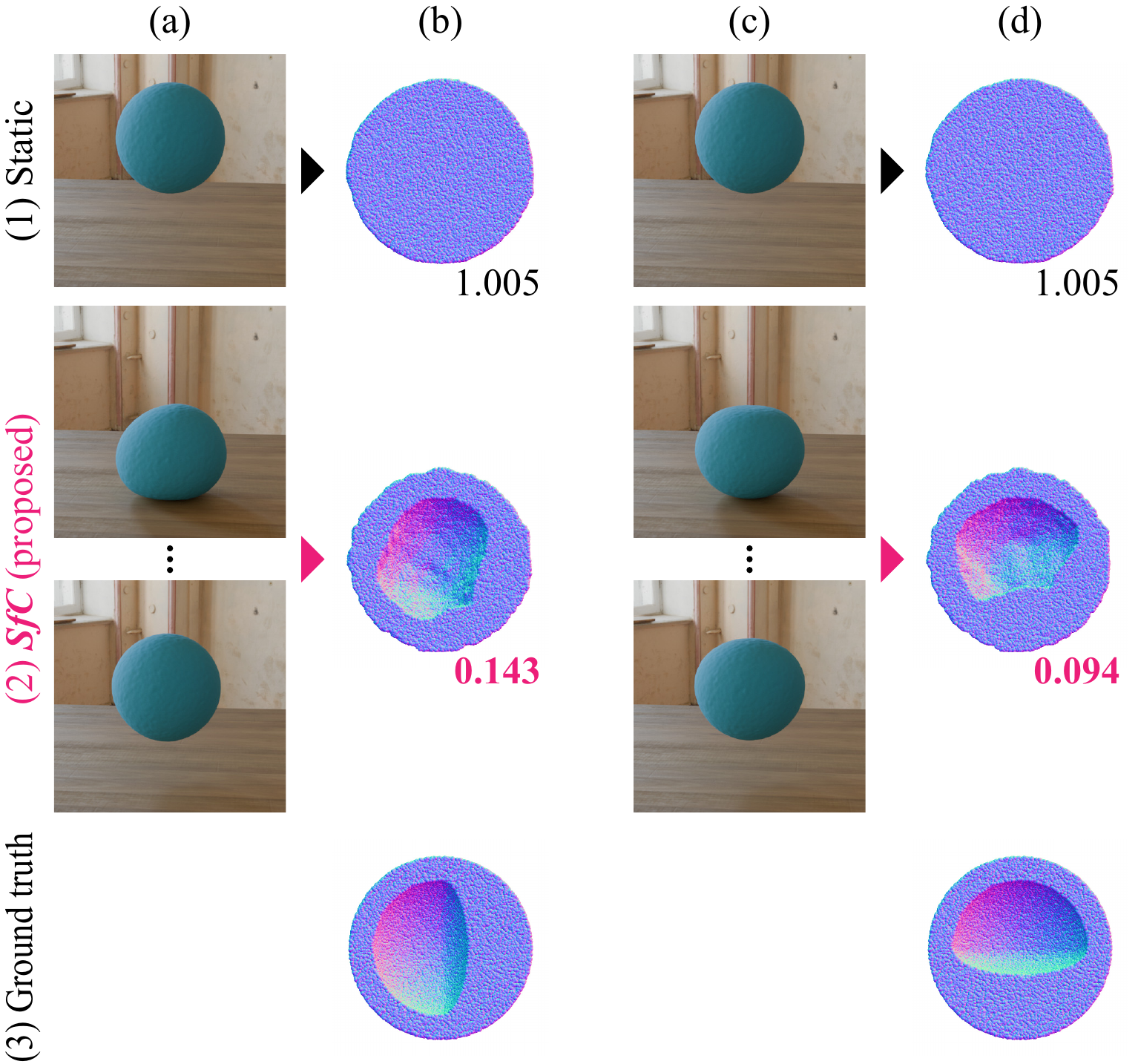}
  \vspace{-5mm}
  \caption{Concept of \textit{Structure from Collision (SfC)}.
    (a) and (c) Examples of training images taken from a certain viewpoint.
    (b) and (d) Cross-sectional views of the internal structures cut perpendicular to the viewpoint.
    The score indicates the chamfer distance ($\times 10^3$$\downarrow$) between the ground-truth and estimated particles (the smaller, the better).
    Here, two objects appear to be identical in static images (1) but actually have different internal structures (3).
    (1) A static 3D representation learning model cannot distinguish the difference in internal structures (b)(d) because there is no difference in appearance in static images (a)(c).
    (2) To overcome this limitation, we address \textit{SfC}.
    As shown in (a) and (c), changes in shape and appearance during collision are influenced by the internal structure.
    We utilize this property to identify the internal structure of the object.
    Although it is still difficult to identify perfectly owing to its ill-posed nature, the proposed method has succeeded in capturing the bias in the location of the holes (b)(d).}
  \vspace{-3mm}
  \label{fig:concept}
\end{figure}

\section{Introduction}
\label{sec:introduction}

Learning 3D representations from multiview images is a fundamental problem in computer vision and graphics, with applications across various domains, including augmented and virtual reality, gaming, robotics, and autonomous driving.
Recent advancements in neural 3D representations, such as neural radiance fields (NeRF)~\cite{BMildenhallECCV2020} and 3D Gaussian splatting (3DGS)~\cite{BKerblTOG2023}, have enabled the accurate estimation of 3D structures from multiview images and yielded impressive results in novel view synthesis.

However, this benefit is limited to the estimation of the \textit{visible external} structure, and it remains difficult to estimate the \textit{invisible internal} structure hidden behind the surface.\footnote{More strictly, when an object is transparent or translucent, it is possible to estimate the internal structure hidden behind the surface using a volume rendering-based 3D representation learning model (e.g., NeRF~\cite{BMildenhallECCV2020}) because it represents appearance on the basis of cumulative volume densities.
However, this effect is limited when an object is not transparent.
  This study aims to identify the internal structure even in the latter case.}
For example, in Figure~\ref{fig:concept}, the two objects have different internal structures, as shown in Figure~\ref{fig:concept}(3)(b) and (3)(d).
However, they are identical in the static images, as shown in Figure~\ref{fig:concept}(1)(a) and (1)(c).
Consequently, a standard static neural 3D representation learning model (e.g., the voxel-based NeRF~\cite{CSunCVPR2022} used in this example) learns the same internal structures (Figure~\ref{fig:concept}(1)(b) and (1)(d)) and ignores the differences in the internal structures.
This misestimation of the internal structure can cause issues in practical applications, such as reproducing and simulating objects in virtual and augmented reality and controlling forces during interactions with objects in robotics.

To overcome this limitation, we address a novel task called \textit{Structure from Collision (SfC)}, the objective of which is to identify the structure (including the invisible internal structures) of an object based on observations at collision.
This is motivated by the observation that changes in appearance and shape during collisions are influenced by the internal structures.
For example, as shown in Figure~\ref{fig:concept}(2)(a) and (2)(c), when a hole exists inside the sphere on the left side (Figure~\ref{fig:concept}(3)(b)) or on the upper side (Figure~\ref{fig:concept}(3)(d)), the sphere crumples when it hits the ground.
We use this property to identify the internal structure of the object.

We formulated \textit{SfC} to optimize the invisible internal structures of an object under \textit{physical}, \textit{appearance (i.e., visible external structure)-preserving}, and \textit{keyframe constraints}.
Specifically, we implemented this approach using \textit{SfC-NeRF}, which consists of four components.

\textit{(1) Physical constraints.}
\textit{SfC} is ill-posed because the observable data represent just one of many possible solutions.
To address this issue, we narrow the solution space by incorporating \textit{physical constraints}, specifically by using physics-augmented continuum NeRF (PAC-NeRF)~\cite{XLiICLR2023}.

\textit{(2) Appearance-preserving constraints.}
Owing to the recent advancements in neural 3D representations, learning \textit{visible external} structures is easier than learning \textit{invisible internal structures}.
Accordingly, we first learn the external structures using a standard static neural 3D representation learning model (voxel-based NeRF~\cite{CSunCVPR2022} in practice) using the first frame (Figure~\ref{fig:concept}(1)).
We then optimize the internal structures using a video sequence (Figure~\ref{fig:concept}(2)).
In the second step, to avoid damaging the external structures learned in the first step when fitting the entire video, we introduce \textit{appearance-preserving constraints} that optimize the internal structures while preserving the external structures.

\textit{(3) Keyframe constraints.}
In a collision video, a specific frame (e.g., immediately after a collision) is effective in explaining the shape change caused by the collision.
Accordingly, we incorporate \textit{keyframe constraints} to strengthen shape learning in the keyframe.

\textit{(4) Volume annealing.}
To avoid becoming stuck in undesirable local optima owing to the existence of multiple solutions, we developed \textit{volume annealing}, in which the global optimum is searched for through an annealing process that repeatedly reduces and expands the volume.

We comprehensively evaluated the proposed method using a dataset containing $115$ objects with diverse structures (i.e., various cavity shapes, locations, and sizes) and material properties.
Our results reveal the properties of \textit{SfC} and demonstrate the effectiveness of \textit{SfC-NeRF}.
Figure~\ref{fig:concept}(2)(b) and (d) show examples of the results obtained using \textit{SfC-NeRF}.
Although it is challenging to perfectly match the internal structures to the ground truth owing to the high degrees of freedom in the solution, \textit{SfC-NeRF} successfully identified the deviation of the hole inside the sphere.

The contributions of this study are threefold:
\begin{itemize}
\item We address a novel task called \textit{SfC}, whose aim is to identify structures (including the internal structures) from the appearance changes at collision.
\item To solve \textit{SfC}, we propose \textit{SfC-NeRF}, which consists of four components: \textit{physical}, \textit{appearance-preserving}, and \textit{keyframe constraints}, and \textit{volume annealing}.
\item Through extensive experiments on $115$ objects, we demonstrate the effectiveness of \textit{SfC-NeRF} while clarifying the properties of \textit{SfC}.
  We also provide detailed results and implementation details in the Appendices.
  Video samples are available at the \href{https://www.kecl.ntt.co.jp/people/kaneko.takuhiro/projects/sfc/}{project page}.
\end{itemize}

\section{Related work}
\label{sec:related_work}

\textbf{Neural 3D representations.}
Learning 3D representations is a fundamental problem in computer vision and graphics.
Recent advancements in neural 3D representations, such as NeRF~\cite{BMildenhallECCV2020} and 3DGS~\cite{BKerblTOG2023}, have lead to significant breakthroughs, and various derivative models have been proposed.
These models can be roughly divided into three categories, based on their objectives.
\textit{(1) Improvement of quality} of rendered images or reconstructed 3D data~\cite{KZhangArXiv2020,JBarronICCV2021,JBarronCVPR2022,TBarronICCV2023,DVerbinCVPR2022,BMildenhallCVPR2022,WHuICCV2023,ZYuCVPR2024,ZYanCVPR2024,ZLiangECCV2024,YJiangCVPR2024,JLiuECCV2024,YLiECCV2024},
\textit{(2) improvement of efficiency}, i.e., speeding up and reducing memory usage in training or inference~\cite{TNeffCVF2021,DLindellCVPR2021,AKurzECCV2022,VSitzmannNeurIPS2021,BAttalCVPR2022,MSuhailCVPR2022,HWangECCV2022,DRebainCVPR2021,CReiserICCV2021,SGarbinICCV2021,PHedmanICCV2021,AYu2021ICCV,SFridovichCVPR2022,THuCVPR2022,LLiuNeurIPS2020,CSunCVPR2022,SWizadwongsaCVPR2021,AChenECCV2022,TMullerTOG2022,TKanekoICCV2023,JLeeCVPR2024,SNiedermayrCVPR2024,YChenECCV2024}, and
\textit{(3) incorporation of other modules or functionalities}, such as generative models~\cite{KSchwarzNeurIPS2020,EChanCVPR2021,MNiemeyerCVPR2021,JGuICLR2022,EChanCVPR2022,YDengCVPR2022,TKanekoCVPR2022,YXueCVPR2022,ISkorokhodovNeurIPS2022,BPooleICLR2023,ERChanICCV2023,CHLinCVPR2023,RChenICCV2023,ZWangNeurIPS2023,RGaoNeurIPS2024,JTangICLR2024,TYiCVPR2024,JTangECCV2024,SZhouECCV2024} and physics/dynamics~\cite{ZLiCVPR2021,APumarolaCVPR2021,GGafniCVPR2021,ETretschkICCV2021,KParkICCV2021,MChuTOG2022,SGuanICML2022,XLiICLR2023,TKanekoCVPR2024,JAbouWACV2024,YFengCVPR2024,ZYangICLR2024,ZYangCVPR2024,JLuiten3DV2024,ZLuCVPR2024,TXieCVPR2024,YJiangTOG2024,RZQiuECCV2024,JAbouCoRL2024}.
This study focuses on the third category, aiming to discover internal structures based on dynamic observations under physical constraints.
Because these models are mutually developed, applying the proposed approach to other models presents an interesting direction for future research.

\smallskip\noindent
\textbf{Dynamic neural 3D representations.}
Dynamic neural 3D representations can be classified into two categories, based on whether they incorporate physics.
\textit{(1) Non- (or weak) physics-informed models}~\cite{ZLiCVPR2021,APumarolaCVPR2021,GGafniCVPR2021,ETretschkICCV2021,KParkICCV2021,ZYangICLR2024,ZYangCVPR2024,JLuiten3DV2024,ZLuCVPR2024} and
\textit{(2) physics-informed models}~\cite{MChuTOG2022,SGuanICML2022,XLiICLR2023,TKanekoCVPR2024,JAbouWACV2024,YFengCVPR2024,TXieCVPR2024,YJiangTOG2024,RZQiuECCV2024,JAbouCoRL2024}.
The first category offers flexibility, and can be applied to scenes or objects that are difficult to describe physically.
However, it requires a large amount of training data and lacks interpretability because of its fully data-driven black-box nature.
By introducing physics, the second category provides a better interpretability and narrows the solution space.
However, they lose flexibility and are difficult to apply to scenes or objects that cannot be explained by physics.
This study adopts a physics-informed model (the second-category strategy) because \textit{SfC} is an ill-posed problem, and physics plays an important role in narrowing the solution space.
However, in the future, it would be interesting to explore how the first-category strategy can be used by expanding data and developing new theories.

\smallskip\noindent
\textbf{Physics-informed neural 3D representations.}
Physics-informed neural 3D representations can be divided into two categories based on the problem setting.
\textit{(1) Forward engineering}~\cite{YFengCVPR2024,TXieCVPR2024,YJiangTOG2024,RZQiuECCV2024}, where a physics-informed model is optimized to fit \textit{static} scenes or objects, and then physics-informed dynamic simulations or interactive manipulations are performed.
In most cases, the inside of the object is assumed to be \textit{filled}, and internal factors, such as physical properties, are \textit{manually adjusted} to achieve visually plausible results.
\textit{(2) Reverse engineering}~\cite{MChuTOG2022,SGuanICML2022,XLiICLR2023,TKanekoCVPR2024,JAbouWACV2024,JAbouCoRL2024}, which focuses on system identification---identifying internal factors (e.g., physical properties) from \textit{dynamic} observations (i.e., video sequences).
This study falls into the second category because it aims to reverse engineer the \textit{internal structure}, which is hidden but essential for describing the system, from collision videos.

Reverse engineering is generally ill-posed because the observable data represent only one of the many possible solutions.
To address this issue, the methods in this category typically impose assumptions on internal factors that are not optimized.
Previous studies have made various assumptions regarding the internal structure, which is the main focus of this study.
For example, \cite{MChuTOG2022} assumes that an object, such as smoke,  is \textit{translucent}, allowing \textit{part of the internal structure to be visible}.
Other studies~\cite{SGuanICML2022,XLiICLR2023,TKanekoCVPR2024,JAbouWACV2024,JAbouCoRL2024} considered \textit{non-transparent} objects but assumeed that the interior is \textit{filled}.
Consequently, \textit{non-transparent and unfilled objects} have not been sufficiently explored.
Therefore, this study focused on such objects.
It is important to note that, as with conventional problems, solving \textit{SfC} is challenging without making any assumptions.
In this study, we assumed that certain internal factors, such as physical properties, are known in advance.
Even with this assumption, as shown in Figure~\ref{fig:concept} (where physical properties, such as mass, Young's modulus, and density, are identical), multiple solutions still exist, making \textit{SfC} a challenging problem.
Details of the problem settings are discussed in Section~\ref{subsec:problem}.

\section{Method}
\label{sec:method}

\subsection{Problem statement}
\label{subsec:problem}

First, we define the \textit{SfC} problem.
Given a set of multiview videos in which objects collide (e.g., Figure~\ref{fig:concept}(2)(a) and (2)(c)), the objective of \textit{SfC} is to identify the structure of the object, including its \textit{invisible internal} structure, based on the appearance changes before and after the collision.
Formally, the training data, i.e., a set of multiview videos, are defined as a collection of ground-truth color observations $\hat{\mathbf{C}}(\mathbf{r}, t)$.
Here, $\mathbf{r} \in \mathbb{R}^3$ is a camera ray defined as $\mathbf{r}(s) = \mathbf{o} + s \mathbf{d}$, where $\mathbf{o} \in \mathbb{R}^3$ is the camera origin, $\mathbf{d} \in \mathbb{S}^2$ is the view direction, and $s \in [s_n, s_f]$ is the distance from $\mathbf{o}$.
During training, $\mathbf{r}$ is sampled from $\hat{\mathcal{R}}$, which is a collection of camera rays in the training dataset.
$t \in \{ t_0, \dots, t_{N - 1} \}$ represents the time, where $N$ is the total number of frames.
Given these data, we aim to estimate the 3D structure (both external and internal ones) of the object $\mathcal{P}^P(t_0)$, which corresponds to the ground truth $\hat{\mathcal{P}}^P(t_0)$.
Here, we represent the 3D structures as particle sets, $\mathcal{P}^P (t_0)$ and $\hat{\mathcal{P}}^P (t_0)$, as shown in Figure~\ref{fig:concept}(b) and (d).
During training, only the external appearance $\hat{\mathbf{C}}(\mathbf{r}, t)$ is observed; $\hat{\mathcal{P}}^P (t_0)$, which includes the internal structure, is not observable.

As discussed in Sections~\ref{sec:introduction} and \ref{sec:related_work}, \textit{SfC} is an ill-posed problem with multiple solutions.
Internal structures and physical properties, such as Young's modulus, have a mutually dependent relationship because both can explain the relationship between strain and stress.
For example, a highly elastic object can be created either by making it hollow or by using soft materials.
To address this issue, PAC-NeRF~\cite{XLiICLR2023} optimizes \textit{physical properties} by assuming that \textit{the inside of the object is filled}.
In contrast, we address a complementary problem, namely optimizing the \textit{internal structure} based on the assumption that the \textit{physical properties are known}.
Specifically, we assume that the physical properties related to the material (e.g., Young's modulus $\hat{E}$, Poisson's ratio $\hat{\nu}$, and density $\hat{\rho}$) and mass $\hat{m}$ are known.
Even with this assumption, \textit{SfC} remains a challenging problem because multiple internal structures can satisfy the same set of physical properties, as shown in Figure~\ref{fig:concept}.

\subsection{Preliminary: PAC-NeRF}
\label{subsec:preliminary}

As explained in the previous subsection, the problem settings differ between the PAC-NeRF study~\cite{XLiICLR2023} and this study.
However, because the proposed model uses PAC-NeRF to describe the physics, we briefly review PAC-NeRF here.
PAC-NeRF is a variant of NeRF that bridges the Eulerian grid-based scene representation~\cite{CSunCVPR2022} with a Lagrangian particle-based differentiable physical simulation~\cite{YHuICLR2020} for continuum materials, such as elastic materials, plasticine, sand, and fluids.
PAC-NeRF obtains this functionality using three components: a continuum NeRF, a particle–grid interconverter, and a Lagrangian field.

\smallskip\noindent
\textbf{Continuum NeRF.}
Continuum NeRF is built on dynamic NeRF (NeRF for a dynamic scene)~\cite{APumarolaCVPR2021}.
In the dynamic NeRF, the volume density and color fields for position $\mathbf{x}$, view direction $\mathbf{d}$, and time $t$ are defined as $\sigma (\mathbf{x}, t)$ and $\mathbf{c} (\mathbf{x}, \mathbf{d}, t)$, respectively.
On this basis, the color of each pixel $\mathbf{C}(\mathbf{r}, t)$ is rendered using volume rendering~\cite{BMildenhallECCV2020}:
\begin{flalign}
  \label{eq:volume_rendering}
  \mathbf{C}(\mathbf{r}, t)
  & = \int_{s_n}^{s_f} T_{\mathbf{r}}(s, t) \sigma (\mathbf{r}(s), t) \mathbf{c} (\mathbf{r}(s), \mathbf{d}, t) ds,
  \\
  T_{\mathbf{r}}(s, t)
  & = \exp \left( - \int_{s_n}^s \sigma (\mathbf{r}(u), t) du \right).
\end{flalign}
This model can be trained using a pixel loss.
\begin{flalign}
  \label{eq:pixel_loss}
  \mathcal{L}_{\text{pixel}}
  = \frac{1}{N} \sum_{i = 0}^{N - 1} \frac{1}{| \hat{\mathcal{R}} |} \sum_{\mathbf{r} \in \hat{\mathcal{R}}}
  \| \mathbf{C}(\mathbf{r}, t_i) - \hat{\mathbf{C}}(\mathbf{r}, t_i) \|_2^2.
\end{flalign}
Dynamic NeRF is extended to continuum NeRF to describe the dynamics of continuum materials.
This is achieved by applying the conservation laws to $\sigma (\mathbf{x}, t)$ and $\mathbf{c} (\mathbf{x}, \mathbf{d}, t)$:
\begin{flalign}
  \label{eq:conservation_law}
  \frac{D\sigma}{Dt} = 0, \:\:
  \frac{D {\mathbf{c}}}{Dt} = \mathbf{0},
\end{flalign}
where $\frac{D \phi}{D t} = \frac{\partial \phi}{\partial t} + \mathbf{v} \cdot \nabla \phi$ for an arbitrary time-dependent field $\phi (\mathbf{x}, t)$.
Here, $\mathbf{v}$ is the velocity field and obeys momentum conservation for continuum materials:
\begin{flalign}
  \label{eq:momentum_conservation}
  \rho \frac{D \mathbf{v}}{Dt} = \nabla \cdot \mathbf{T} + \rho \mathbf{g},
\end{flalign}
where $\rho$ is the physical density field, $\mathbf{T}$ is the internal Cauchy stress tensor, and $\mathbf{g}$ is the gravitational acceleration.
This equation can be solved differentially using the differentiable material point method (DiffMPM)~\cite{YHuICLR2020}.

\smallskip\noindent
\textbf{Particle--grid interconverter.}
DiffMPM is a particle-based method that conducts simulations in a Lagrangian space.
However, these particles do not necessarily lie on the ray, which makes rendering difficult.
Considering this, PAC-NeRF renders in an Eulerian grid space with voxel-based NeRF~\cite{CSunCVPR2022} and bridges these two spaces using grid-to-particle (G2P) and particle-to-grid (P2G) conversions:
\begin{flalign}
  \label{eq:g2p_p2g}
  \mathcal{F}_p^P \approx \sum_i w_{ip} \mathcal{F}_i^G, \:\:
  \mathcal{F}_i^G \approx \frac{\sum_p w_{ip} \mathcal{F}_p^P}{\sum_p w_{ip}},
\end{flalign}
where $\mathcal{F}^X = \{ \sigma^X (\mathbf{x}, t), \mathbf{c}^X (\mathbf{x}, \mathbf{d}, t) \}$ for $X \in \{ G, P \}$.
Here, $G$ and $P$ represent the Eulerian and Lagrangian views, respectively.
When $\mathcal{F}^X$ is used with a subscript, that is, $\mathcal{F}_x^X$ ($x \in \{ i, p \}$), the subscripts $i$ and $p$ indicate the grid node and particle index, respectively.
$w_{ip}$ denotes the weight of the trilinear shape function defined at $i$ and evaluated at $p$.

\smallskip\noindent
\textbf{Lagrangian field.}
The physical simulation and rendering pipeline in PAC-NeRF proceeds as follows:
(1) Volume densities and colors are initialized over the first frame of the video sequence in an Eulerian grid field, $\mathcal{F}^{G'} (t_0)$.
Here, we use the superscript $G'$ to distinguish $\mathcal{F}^{G'}$ from $\mathcal{F}^{G}$ used in Step (4).
(2) Using the G2P process, $\mathcal{F}^{G'} (t_0)$ is converted into a Lagrangian particle field, $\mathcal{F}^{P} (t_0)$.
In this step, particles $\mathcal{P}^P (t_0)$ are sampled at intervals of half the grid, that is, $\frac{\Delta x}{2}$ (where $\Delta x$ is the grid size), with random fluctuations.
The alpha value (or amount of opacity) $\alpha_p^P$ is calculated for each particle using $\alpha_p^P = 1 - \exp(-\text{softplus}(\sigma_p^P))$, and a particle is removed if $\mathcal{\alpha}_p^P < \epsilon$ ($\epsilon = 10^{-3}$ in practice).
(3) The particle field in the next step, $\mathcal{F}^{P} (t_1)$, is calculated from $\mathcal{F}^{P} (t_0)$ using DiffMPM~\cite{YHuICLR2020}, where $t_1 = t_0 + \delta t$, and $\delta t$ is the duration of the time step.
Similarly, the particle field at $t$, $\mathcal{F}^P (t)$, is calculated for $t \in \{ t_0, \dots, t_{N-1} \}$.
(4) Using the P2G process, $\mathcal{F}^P (t)$ is converted into an Eulerian grid field, $\mathcal{F}^G (t)$.
(5) $\mathbf{C}(\mathbf{r}, t)$ is rendered based on $\mathcal{F}^G (t)$ by using voxel-based volume rendering~\cite{CSunCVPR2022}.

During training, two-step optimization is conducted.
(i) $\mathcal{F}^{G'} (t_0)$ is initially optimized using the first frame of the video sequence by conducting processes (1)--(5) for $t = t_0$.
(ii) Physical properties, such as the Young's modulus $E$ and  Poisson's ratio $\nu$, are optimized for the entire video sequence by conducting processes (1)--(5) for $t \in \{ t_0, \dots, t_{N - 1} \}$.
In both optimizations, $\mathcal{L}_{\text{pixel}}$ (Equation~\ref{eq:pixel_loss}) is used as the objective function.

\begin{figure*}[t]
  \begin{center}
    \includegraphics[width=\textwidth]{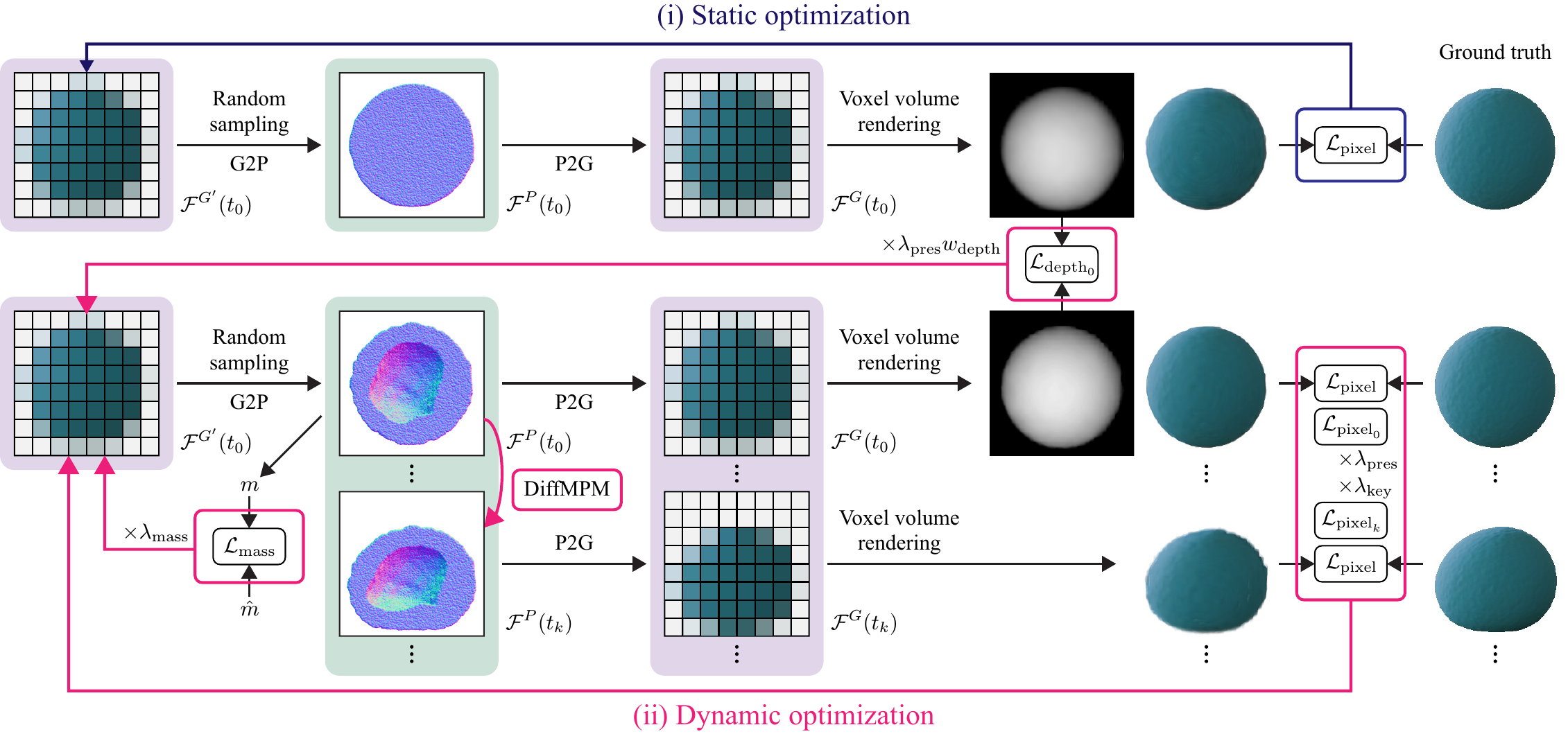}
  \end{center}
  \vspace{-5mm}
  \caption{Optimization pipelines of SfC-NeRF.
    (i) The grid field $\mathcal{F}^{G'}(t_0)$ is initially optimized using the first frame of the video sequence.
    (ii) Subsequently, the structure (i.e., volume density $\sigma^{G'}(t_0) \in \mathcal{F}^{G'}(t_0)$) of the object is optimized through the entire video sequence with physical constraints ($\mathcal{L}_{\text{mass}}$ and DiffMPM), appearance-preserving constraints (i.e., $\mathcal{L}_{\text{pixel}_0}$ and $\mathcal{L}_{\text{depth}_0}$), and keyframe constraints ($\mathcal{L}_{\text{pixel}_k}$) along with a standard pixel loss ($\mathcal{L}_{\text{pixel}}$).}
  \vspace{-3mm}
  \label{fig:pipelines}
\end{figure*}

\subsection{Proposal: SfC-NeRF}
\label{subsec:proposal}

Similar to PAC-NeRF, \textit{SfC-NeRF} performs two-step optimization, as shown in Figure~\ref{fig:pipelines}.
The first-step optimization (Figure~\ref{fig:pipelines}(i)) is the same as that in PAC-NeRF, that is, $\mathcal{F}^{G'} (t_0)$ is initially optimized using the first frame of the video sequence.
In this step, the filled object is learned, as shown in Figure~\ref{fig:concept}(1).
In contrast, the second step of the optimization (Figure~\ref{fig:pipelines}(ii)) differs because of the difference in the optimization target.
In the PAC-NeRF, the \textit{physical properties} are optimized in this step, whereas in the \textit{SfC-NeRF}, the \textit{internal structure} is optimized.
Specifically, as explained in the previous section, we obtain particles $\mathcal{P}^P (t_0)$ based on $\sigma^P(t_0)$, which is calculated from $\sigma^{G'}(t_0)$ (Steps (1) and (2)).
Therefore, we select $\sigma^{G'} (t_0)$ as the optimization target.\footnote{Note that Lagrangian particle optimization (LPO)~\cite{TKanekoCVPR2024} also considers a similar optimization (i.e., optimizing $\mathcal{F}^P (t_0)$ or $\mathcal{F}^{G'} (t_0)$ through a video sequence) for few-shot (sparse view) learning.
  However, it aims to compensate for the \textit{external} structure where the viewpoint is missing and \textit{has not} sufficiently considered the components necessary for estimating the internal structures, which are discussed in the following paragraphs.
  We demonstrate the limitations of LPO in our experiments (Section~\ref{sec:experiments}).}
In particular, we formulate \textit{SfC} as a problem of optimizing $\sigma^{G'}(t_0)$ under \textit{physical}, \textit{appearance (i.e., external structure)-preserving}, and \textit{keyframe constraints}, along with \textit{volume annealing}.

\smallskip\noindent
\textbf{Physical constraints.}
As discussed in Section~\ref{subsec:problem}, we assume that the physical properties related to the material (e.g., Young's modulus $\hat{E}$, Poisson's ratio $\hat{\nu}$, and density $\hat{\rho}$) and mass $\hat{m}$ are known.
We utilize them to narrow the solution space of \textit{SfC}.

\smallskip\noindent
\textit{Physical constraints on material properties.}
We can reflect material-specific physical properties (e.g., Young's modulus $\hat{E}$, Poisson's ratio $\hat{\nu}$, and density $\hat{\rho}$) explicitly when constructing DiffMPM~\cite{YHuICLR2020}.
Motivated by this fact, we optimize $\sigma^{G'}(t_0)$ under the \textit{explicit material-specific physical constraints imposed by DiffMPM}.

\smallskip\noindent
\textit{Physical constraints on mass.}
Unlike physical material properties, mass is not determined only by the material and varies depending on the individual objects.
Therefore, instead of explicitly representing the mass in DiffMPM, we constrain the mass using a \textit{mass loss}.
\begin{flalign}
  \label{eq:mass_loss}
  \mathcal{L}_{\text{mass}} & = \| \log_{10}(m) - \log_{10}(\hat{m}) \|_2^2,
  \\
  \label{eq:mass}
  m & = \sum_{p \in \mathcal{P}^P(t_0)} \hat{\rho} \cdot \left( \frac{\Delta x}{2} \right)^3 \cdot \alpha_p^P,
\end{flalign}
where $m$ and $\hat{m}$ are the estimated and ground-truth masses, respectively.
In Equation~\ref{eq:mass}, $m$ is computed by summarizing the mass of each particle indexed by $p \in \mathcal{P}^P(t_0)$, where the mass of each particle is given by the product of the physical density $\hat{\rho}$, the unit volume of a particle $\left(\frac{\Delta x}{2} \right)^3$, and the alpha value $\alpha_p^P$.
In Equation~\ref{eq:mass_loss}, we employ a logarithmic scale to prioritize scale matching.

\smallskip\noindent
\textbf{Appearance-preserving constraints.}
As mentioned above, we use two-step optimization:
(i) $\mathcal{F}^{G'}$ is initially optimized using the first frame of the video sequence (Figure~\ref{fig:pipelines}(i)).
(ii) $\sigma^{G'}$ is optimized through a video sequence (Figure~\ref{fig:pipelines}(ii)).
In Step (ii), the external structure (or surface) learned in Step (i) does not need to be changed, considering that learning the external structure is easier than learning the internal structure.
However, the physical constraints discussed above are not sufficient to satisfy this requirement.
Hence, we introduced appearance-preserving constraints at both the loss and training scheme levels.

\smallskip\noindent
\textit{Appearance-preserving loss.}
The standard pixel loss (Equation~\ref{eq:pixel_loss}) treats the loss for each frame equally.
This is insufficient to prevent the external structure, which is well-learned in Step (i), from changing as a result of the fitting of the entire video sequence.
Hence, we employ a \textit{pixel-preserving loss} that preserves the appearance of the initial frame.
\begin{flalign}
  \label{eq:pixel_pres_loss}
  \mathcal{L}_{\text{pixel}_0}
  = \frac{1}{| \hat{\mathcal{R}} |} \sum_{\mathbf{r} \in \hat{\mathcal{R}}}
  \| \mathbf{C}(\mathbf{r}, t_0) - \hat{\mathbf{C}}(\mathbf{r}, t_0) \|_2^2.
\end{flalign}
This is a variant of pixel loss (Equation~\ref{eq:pixel_loss}) when $N = 1$.
Because the constraints on the 2D projection plane alone are insufficient for preserving the 3D structure (e.g., objects with reversed concavity may be learned), we also incorporate a \textit{depth-preserving loss} to encourage the preservation of the depth of the initial frame.
\begin{flalign}
  \label{eq:depth_pres_loss}
  \mathcal{L}_{\text{depth}_0}
  & = \frac{1}{| \hat{\mathcal{R}} |} \sum_{\mathbf{r} \in \hat{\mathcal{R}}}
    ( \| \Delta_h Z(\mathbf{r}, t_0) - \Delta_h \tilde{Z}(\mathbf{r}, t_0) \|_2^2,
  \nonumber \\
  & \:\:\:\:\:\:\:\:\:\:\:\:\:\:\:\:\:\:
    + \| \Delta_v Z(\mathbf{r}, t_0) - \Delta_v \tilde{Z}(\mathbf{r}, t_0) \|_2^2),
\end{flalign}
where $Z(\mathbf{r}, t_0)$ and $\tilde{Z}(\mathbf{r}, t_0)$ are the depths predicted by the current model and the model before performing Step (ii), respectively.
We use $\tilde{Z}(\mathbf{r}, t_0)$ because the ground-truth depth is not observable.
$Z(\mathbf{r}, t_0)$ is calculated by $Z(\mathbf{r}, t_0) = \int_{s_n}^{s_f} T_{\mathbf{r}}(s, t) \sigma (\mathbf{r}(s), t) s ds$, and $\tilde{Z}(\mathbf{r}, t_0)$ is calculated in a similar manner.
$\Delta_h$ and $\Delta_v$ are operations that calculate the horizontal and vertical differences between adjacent pixels, respectively.
We compare the differences rather than the raw data to mitigate the negative effects of depth estimation errors.

\smallskip\noindent
\textit{Appearance-preserving training.}
Ideally, when an object is non-transparent, its appearance is not expected to change, even if the internal volume density is changed.
However, in preliminary experiments, we found that it is difficult to retain the appearance learned in Step (i) through a simple adaptation of the appearance-preserving losses.
This motivated us to employ \textit{appearance-preserving training}, that is, reoptimizing $\mathcal{F}^{G'}(t_0)$ using the first frames of the video sequence every time after optimizing $\sigma_{G'}(t_0)$ for the entire video sequence.

\smallskip\noindent
\textbf{Keyframe constraints.}
As mentioned in the explanation of appearance-preserving loss, the standard pixel loss treats the loss for each frame equally.
However, in preliminary experiments, we found that certain frames, particularly the frame immediately after the collision, were useful for explaining shape changes due to the internal structures.
Based on this observation, we impose a \textit{keyframe pixel loss} defined as follows:
\begin{flalign}
  \label{eq:key_frame_loss}
  \mathcal{L}_{\text{pixel}_k} = \frac{1}{| \hat{\mathcal{R}} |} \sum_{\mathbf{r} \in \hat{\mathcal{R}}}
  \| \mathbf{C}(\mathbf{r}, t_k) - \hat{\mathbf{C}}(\mathbf{r}, t_k) \|_2^2,
\end{flalign}
where $k$ is the keyframe index (the frame immediately after the collision is used in practice).

\smallskip\noindent
\textbf{Volume annealing.}
As discussed previously, we begin the optimization from the state in which the interior of the object is filled (Figure~\ref{fig:pipelines}(i)).
The internal structure is then optimized by reducing the volume using the aforementioned techniques (Figure~\ref{fig:pipelines}(ii)).
Owing to these learning dynamics, if the volume reduction goes in the wrong direction and leads to a local optimum, it becomes challenging to determine the global optimum.
To address this issue, we introduce \textit{volume annealing}, which involves alternating between the volume reduction and expansion.
This strategy facilitates the search for a global optimum.
Specifically, we implement the volume expansion by successively performing the G2P and P2G processes and replacing the obtained $\mathcal{F}^G(t_0)$ with $\mathcal{F}^{G'}(t_0)$.

\smallskip\noindent
\textbf{Full objective.}
The full objective used in Step (ii) is expressed as follows:
\begin{flalign}
  \label{eq:full}
  & \mathcal{L}_{\text{full}}
  = \mathcal{L}_{\text{pixel}}
  + \lambda_{\text{mass}} \mathcal{L}_{\text{mass}}
  \nonumber\\
  & \:\:\:
    + \lambda_{\text{pres}} (\mathcal{L}_{\text{pixel}_0} + w_{\text{depth}} \mathcal{L}_{\text{depth}_0})
  + \lambda_{\text{key}} \mathcal{L}_{\text{pixel}_k}
\end{flalign}
where $\lambda_{\text{mass}}$, $\lambda_{\text{pres}}$, $w_{\text{depth}}$, and $\lambda_{\text{key}}$ are the weighting hyperparameters.
The effect of each loss is analyzed using the ablation study presented in Section~\ref{sec:experiments}.

\section{Experiments}
\label{sec:experiments}

\subsection{Experimental setup}
\label{subsec:experimental_setup}

We conducted three experiments to evaluate \textit{SfC-NeRF} and explore the properties of \textit{SfC}.
First, we examined the impact of changes in the internal structure, focusing on the \textit{cavity sizes} (Experiment I in Section~\ref{subsec:experiment1}) and \textit{locations} (Experiment II in Section~\ref{subsec:experiment2}).
We then explored the effect of the \textit{material properties} in Experiment III (Section~\ref{subsec:experiment3}).
The main results are summarized here, with the detailed results and implementation details provided in the Appendices.
Video samples are available at the \href{https://www.kecl.ntt.co.jp/people/kaneko.takuhiro/projects/sfc/}{project page}.

\begin{figure}[t]
  \centering
  \includegraphics[width=\linewidth]{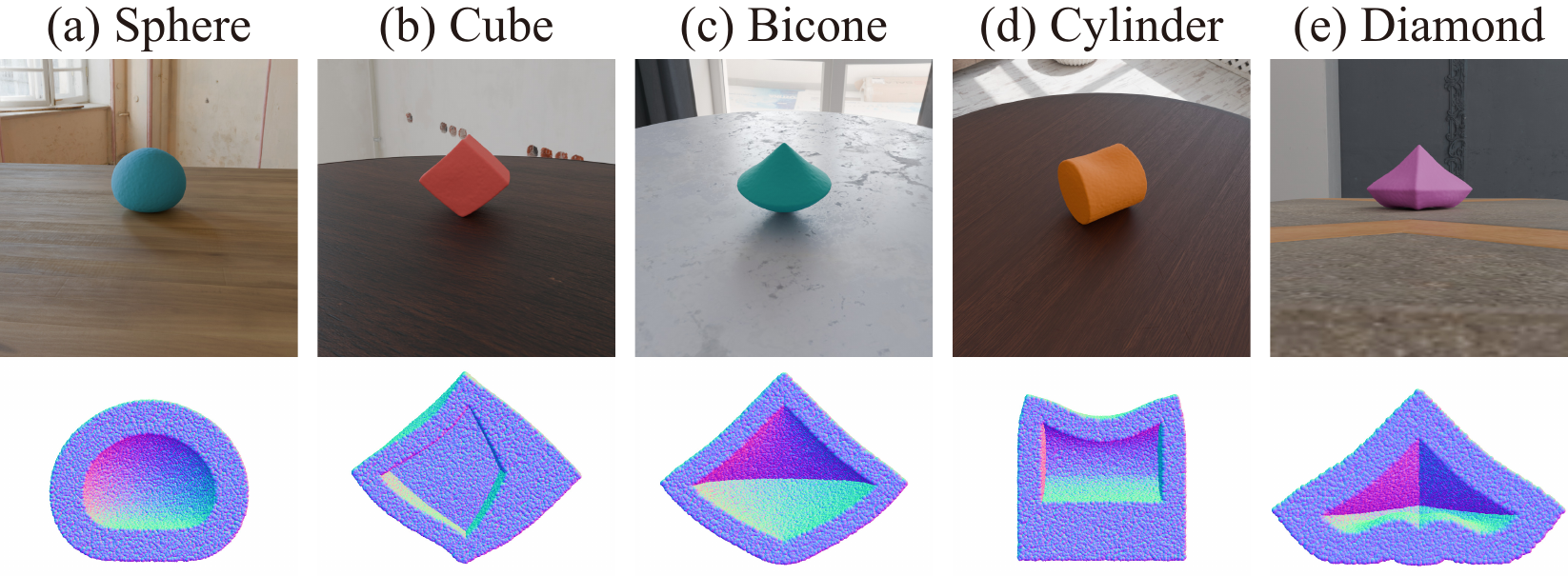}
  \vspace{-6mm}
  \caption{Examples of the data in the SfC dataset.}
  \vspace{-5mm}
  \label{fig:dataset}
\end{figure}

\smallskip\noindent
\textbf{Dataset.}
Because \textit{SfC} is a new task and there is no established dataset, we created a new dataset called the \textit{SfC dataset} based on the protocol of the PAC-NeRF study~\cite{XLiICLR2023}.
We prepared $115$ objects by changing their external shapes, internal structures, and materials.
Figure~\ref{fig:dataset} shows examples of the data in this dataset.
First, we prepared five external shapes: \textit{sphere}, \textit{cube}, \textit{bicone}, \textit{cylinder}, and \textit{diamond}.
Regarding the internal structure and material, we set the default values as follows:
the cavity size rate for filled object, $s_c$, was set to $(\frac{2}{3})^3$,
the cavity location, $l_c$, was set at the center, and
the material was defined as an elastic material with Young's modulus $\hat{E} = 10^6$ and Poisson's ratio $\hat{\nu} = 0.3$.
Under these default properties, one of them was changed as follows:
(a) Three differently sized cavities: $s_c \in \{ 0, (\frac{1}{2})^3, (\frac{3}{4})^3 \}$.
(b) Four different cavity locations: center $l_c$ is moved $\{ \text{up}, \text{down}, \text{left}, \text{right} \}$.
(c) Eight different elastic materials: those with four different Young's moduli $\hat{E} \in \{ 2.5 \times 10^5, 5 \times 10^5, 2 \times 10^6, 4 \times 10^6\}$ and four different Poisson's ratios $\hat{\nu} \in \{ 0.2, 0.25, 0.35, 0.4 \}$.
Seven different materials: two Newtonian fluids, two non-Newtonian fluids, two plasticines, and one sand.
Their physical properties were derived from the PAC-NeRF dataset~\cite{XLiICLR2023}.
Thus, we created $5$ external shapes $\times$ ($1$ default $+$ $3$ sizes $+$ $4$ locations $+$ $(8 + 7)$ materials) $= 115$ objects.

Following the PAC-NeRF study~\cite{XLiICLR2023}, ground-truth data were generated using the MLS-MPM simulator~\cite{YHuTOG2018}, where each object fell freely under the influence of gravity and collided with the ground plane.
Images were rendered under various environmental lighting conditions and ground textures using a photorealistic renderer.
Each scene was captured from 11 viewpoints, including an object, using cameras spaced in the upper hemisphere.

\smallskip\noindent
\textbf{Preprocessing.}
Following the PAC-NeRF study~\cite{XLiICLR2023}, we made two assumptions and performed preprocessing to focus on solving \textit{SfC}.
(1) The intrinsic and extrinsic parameters of the cameras are known.
(2) Collision objects, such as the ground plane, are known.
As mentioned in~\cite{XLiICLR2023}, the latter can be easily estimated from observed images.
For preprocessing, we applied video matting~\cite{SLinCVPR2021} to exclude static background objects, and concentrated the computation on the object of interest.
This process provides a background segmentation mask $\hat{B}(\mathbf{r}, t)$.
NeRF can estimate a background segmentation mask $B(\mathbf{r}, t)$ using $B(\mathbf{r}, t) = 1 - T_{\mathbf{r}}(s_f, t)$.
Taking advantage of this property, we also used a background loss $\mathcal{L}_{\text{bg}} = \| B(\mathbf{r}, t) - \hat{B}(\mathbf{r}, t) \|_2^2$ when calculating the pixel-related losses ($\mathcal{L}_{\text{pixel}}$, $\mathcal{L}_{\text{pixel}_0}$, and $\mathcal{L}_{\text{pixel}_k}$) with a weighting hyperparameter of $w_{\text{bg}}$.
In the experiments, this technique was applied to all the models.

\smallskip\noindent
\textbf{Comparison models.}
Because there is no established method for \textit{SfC}, we adapted previous methods to make them suitable for \textit{SfC}.
Specifically, we used grid optimization (\textit{GO}) and Lagrangian particle optimization (\textit{LPO})~\cite{TKanekoCVPR2024} as baselines.
GO and LPO are improved variants of PAC-NeRF that optimize $\mathcal{F}^{G'}(t_0)$ and $\mathcal{F}^P(t_0)$, respectively, using $\mathcal{L}_{\text{pixel}}$ across a video sequence for few-shot learning.
For a fair comparison with \textit{SfC-NeRF}, GO and LPO were trained using the ground-truth physical properties.
Although the original GO and LPO do not use the mass information for training, it may not be fair to apply it solely to the proposed method.
Therefore, we also examined \textit{GO$_{\text{mass}}$} and \textit{LPO$_{\text{mass}}$}, extensions of GO and LPO that incorporate $\mathcal{L}_{\text{mass}}$.
Furthermore, as an ablation study, we compared \textit{SfC-NeRF} with various variants: \textit{SfC-NeRF$_{-\text{mass}}$}, \textit{SfC-NeRF$_{-\text{APL}}$}, \textit{SfC-NeRF$_{-\text{APT}}$}, \textit{SfC-NeRF$_{-\text{key}}$}, and \textit{SfC-NeRF$_{-\text{VA}}$}, in which the mass loss ($\mathcal{L}_{\text{mass}}$),\footnote{As explained in Appendix~\ref{subsec:training}, the mass information is not only used in the loss but also in adjusting the learning rate.
  In this experiment, we ablated both to simulate a case in which the mass is unknown.} appearance-preserving losses ($\mathcal{L}_{\text{pixel}_0}$ and $\mathcal{L}_{\text{depth}_0}$), appearance-preserving training, keyframe loss ($\mathcal{L}_{\text{pixel}_k}$), and volume annealing were ablated, respectively.
We also examined \textit{Stacic}, a model trained using only the first frame of a video sequence, to assess the effect of optimization across videos.

\smallskip\noindent
\textbf{Evaluation metric.}
As mentioned in Section~\ref{subsec:problem}, we use particles $\mathcal{P}^P (t_0)$ to represent the structure (including the internal structure) of an object and estimate $\mathcal{P}^P (t_0)$ that matches the ground truth $\hat{\mathcal{P}}^P (t_0)$.
Therefore, we evaluate the model by measuring the distance between $\mathcal{P}^P (t_0)$ and $\hat{\mathcal{P}}^P (t_0)$ using the \textit{chamfer distance (CD)}.
The smaller the value, the higher is the degree of matching.

\begin{figure}[t]
  \centering
  \includegraphics[width=\linewidth]{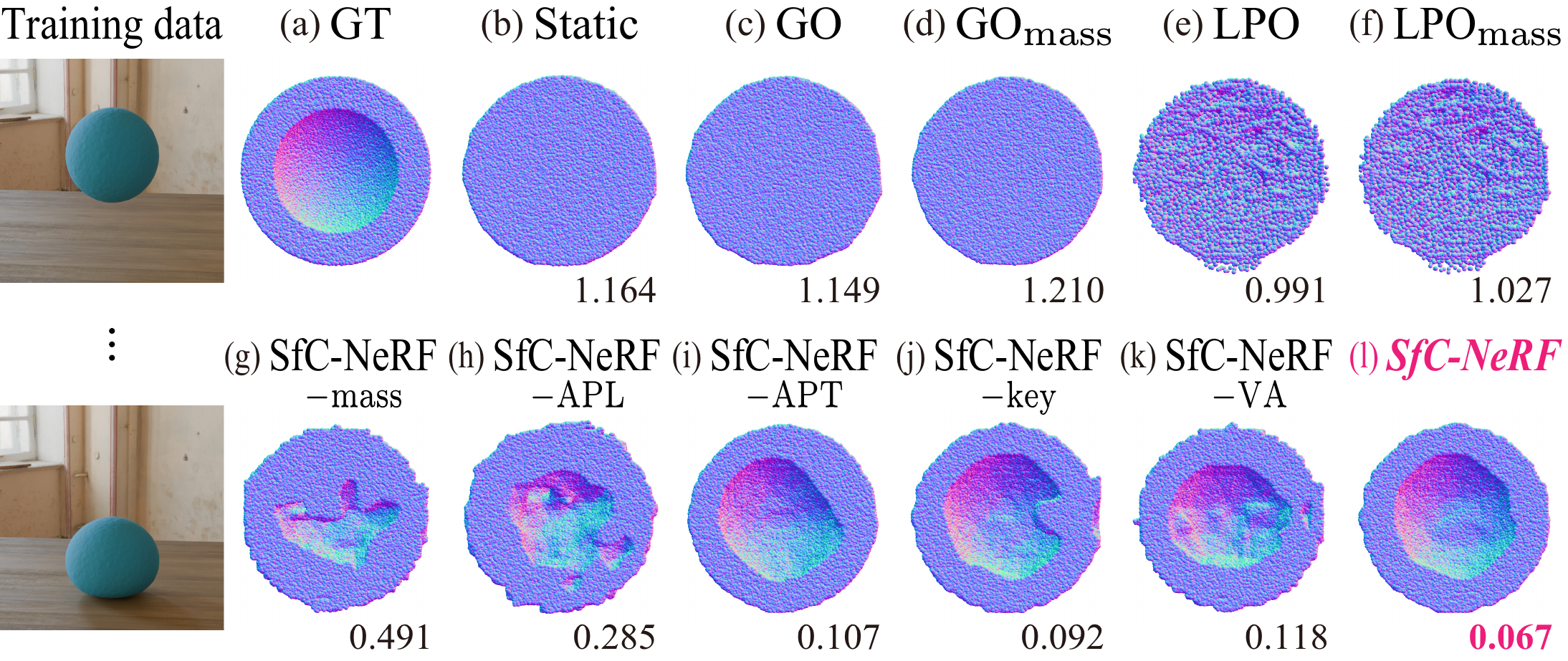}
  \vspace{-7mm}
  \caption{Comparison of learned structures for sphere objects with $s_c = (\frac{2}{3})^3$.
    The score under particles indicates the CD ($\times 10^3$$\downarrow$).
    (c)--(f) GO/LPO failed to determine optimal learning directions.
    (g)--(k) The ablated models failed to avoid improper solutions.
    (l) The full model overcomes these issues and achieves the best CD.}
  \vspace{-5mm}
  \label{fig:results_size}
\end{figure}

\begin{table}[t]
  \centering
  \footnotesize
  \setlength{\tabcolsep}{0.1pt}
  \begin{tabularx}{\columnwidth}{lCCCCC}
    \toprule
    \multicolumn{1}{c}{$s_c$} & $0$ & $(\frac{1}{2})^3$ & $(\frac{2}{3})^3$ & $(\frac{3}{4})^3$ & Avg.
    \\ \midrule
    Static
    & 0.093
    & 0.294
    & 0.920
    & 1.574
    & 0.720
    \\ \midrule
    GO
    & 0.091
    & 0.301
    & 0.941
    & 1.586
    & 0.730
    \\
    GO$_{\text{mass}}$
    & \first{0.081}
    & 0.319
    & 1.244
    & 2.291
    & 0.984
    \\
    LPO
    & 0.092
    & 0.284
    & 0.841
    & 1.406
    & 0.656
    \\
    LPO$_{\text{mass}}$
    & 0.087
    & 0.284
    & 0.876
    & 1.477
    & 0.681
    \\ \midrule
    SfC-NeRF$_{-\text{mass}}$
    & 0.089
    & \third{0.226}
    & 0.550
    & 1.148
    & 0.503
    \\
    SfC-NeRF$_{-\text{APL}}$
    & 0.106
    & 0.423
    & 0.898
    & 1.326
    & 0.688
    \\
    SfC-NeRF$_{-\text{APT}}$
    & \third{0.085}
    & 0.261
    & \third{0.332}
    & 0.661
    & 0.335
    \\
    SfC-NeRF$_{-\text{key}}$
    & \second{0.082}
    & \second{0.127}
    & \second{0.211}
    & \second{0.325}
    & \second{0.186}
    \\
    SfC-NeRF$_{-\text{VA}}$
    & 0.146
    & 0.293
    & 0.370
    & \third{0.456}
    & \third{0.316}
    \\ \midrule
    \textstrong{SfC-NeRF}
    & \first{0.081}
    & \first{0.122}
    & \first{0.195}
    & \first{0.262}
    & \first{0.165}
    \\ \bottomrule
  \end{tabularx}
  \vspace{-3mm}
  \caption{Comparison of CD ($\times 10^3$$\downarrow$) when varying the cavity size $s_c$.
    The scores were averaged over five external shapes.}
  \vspace{-2mm}
  \label{tab:size}
\end{table}

\subsection{Experiment I: Influence of cavity size}
\label{subsec:experiment1}

First, we investigated the influence of the \textit{cavity size} inside the object.
Table~\ref{tab:size} summarizes the quantitative results, and the qualitative results are presented in Figure~\ref{fig:results_size}, Appendix~\ref{sec:qualitative_results_experiment1_2}, and the \href{https://www.kecl.ntt.co.jp/people/kaneko.takuhiro/projects/sfc/}{project page}.
Our findings are threefold.
\textit{(1) Limitations of GO and LPO~\cite{TKanekoCVPR2024}}.
GO, a simple voxel grid optimization using $\mathcal{L}_{\text{pixel}}$, failed to determine an appropriate optimization direction, which led to the deterioration of $\mathcal{P}^P(t_0)$ as it fits the video.
LPO showed a slight improvement by moving particles within physical constraints via DiffMPM.
However, its effectiveness was limited because significant particle movement could alter the unit volume density, making it difficult to find the optimal internal structure.
Furthermore, in both GO and LPO, using mass knowledge with $\mathcal{L}_{\text{mass}}$ did not improve the performance, possibly because they lacked appearance-preserving mechanisms, and forcing $m$ close to $\hat{m}$ can damage the overall structure.
\textit{(2) Effectiveness of each component.}
The ablation study confirms the importance of each model component.
\textit{(3) Increased difficulty with increased cavity size.}
Because optimization begins in the filled state, large cavity sizes require significant volume changes.
We believe that this is the key reason for the deterioration in performance as the cavity size increases.

\begin{table}[t]
  \centering
  \footnotesize
  \setlength{\tabcolsep}{0.1pt}
  \begin{tabularx}{\columnwidth}{lCCCCC}
    \toprule
    \multicolumn{1}{c}{$l_c$} & left & right & up & down & Avg.
    \\ \midrule
    Static
    & 0.841
    & 0.842
    & 0.815
    & 0.813
    & 0.828
    \\ \midrule
    GO
    & 0.874
    & 0.853
    & 0.878
    & 0.870
    & 0.869
    \\
    GO$_{\text{mass}}$
    & 1.349
    & 1.334
    & 1.104
    & 1.001
    & 1.197
    \\
    LPO
    & 0.791
    & 0.787
    & 0.796
    & 0.743
    & 0.779
    \\
    LPO$_{\text{mass}}$
    & 0.824
    & 0.817
    & 0.828
    & 0.775
    & 0.811
    \\ \midrule
    SfC-NeRF$_{-\text{mass}}$
    & \third{0.513}
    & 0.485
    & 0.705
    & 0.479
    & 0.545
    \\
    SfC-NeRF$_{-\text{APL}}$
    & 0.845
    & 0.783
    & 0.805
    & 0.583
    & 0.754
    \\
    SfC-NeRF$_{-\text{APT}}$
    & 0.624
    & \third{0.428}
    & 0.384
    & 0.464
    & 0.475
    \\
    SfC-NeRF$_{-\text{key}}$
    & \second{0.308}
    & \second{0.296}
    & \second{0.307}
    & \second{0.313}
    & \second{0.306}
    \\
    SfC-NeRF$_{-\text{VA}}$
    & 0.542
    & 0.596
    & \third{0.333}
    & \third{0.385}
    & \third{0.464}
    \\ \midrule
    \multirow{2}{*}{\textstrong{SfC-NeRF}}
    & \first{0.303}
    & \first{0.258}
    & \first{0.274}
    & \first{0.291}
    & \first{0.281}
    \\
    & \gray{(0.367)}
    & \gray{(0.431)}
    & \gray{(0.448)}
    & \gray{(0.417)}
    & \gray{(0.416)}
    \\ \bottomrule
  \end{tabularx}
  \vspace{-3mm}
  \caption{Comparison of CD ($\times 10^3$$\downarrow$) when varying the cavity location $l_c$.
    The gray score in parentheses indicates ACD ($\times 10^3$).}
  \vspace{-4mm}
  \label{tab:location}
\end{table}

\subsection{Experiment II: Influence of cavity location}
\label{subsec:experiment2}

Next, we examined the influence of the \textit{cavity location}.
Table~\ref{tab:location} summarizes the quantitative results, and the qualitative results are presented in Appendix~\ref{sec:qualitative_results_experiment1_2} and \href{https://www.kecl.ntt.co.jp/people/kaneko.takuhiro/projects/sfc/}{project page}.
Similar to Experiment I, we observed two main findings:
\textit{(1) Limitations of GO and LPO.} \textit{(2) Effectiveness of each component}.
In addition, we discuss \textit{(3) how well SfC-NeRF captured the cavity location}.
A simple CD is insufficient for this evaluation because it does not account for the deviations.
Therefore, we calculated the \textit{anti-chamfer distance (ACD)}, which measures the chamfer distance between the predicted particles $\mathcal{P}^P(t_0)$ and the ground-truth particles $\tilde{\mathcal{P}}^{P}(t_0)$, where the cavity is placed on the opposite side.
This distance is expected to be longer than the original CD.
The results confirm that the original CD is smaller than the ACD.
These findings suggests that \textit{SfC-NeRF} can capture the positional deviation of a cavity.

\subsection{Experiment III: Influence of material}
\label{subsec:experiment3}

Finally, we investigated the influence of the \textit{material properties}.
Table~\ref{tab:young_poisson} summarizes the quantitative results for elastic materials when $\hat{E}$ and $\hat{\nu}$ were varied.
Table~\ref{tab:material} summarizes the quantitative results for other materials.
Appendix~\ref{subsec:qualitative_results_experiment3} and \href{https://www.kecl.ntt.co.jp/people/kaneko.takuhiro/projects/sfc/}{project page} present the qualitative results.
These results demonstrate that \textit{SfC-NeRF} improves the structure estimation compared with the initial state, regardless of the material.
However, the rate of improvement depends on the material used.
For example, when an object is soft, its shape changes significantly, making it difficult to capture dynamic changes.
In contrast, when the object is hard, there are fewer shape changes that provide limited cues for estimating the internal structure, making learning more difficult.
Thus, the proposed method is most effective when the object is moderately soft or hard.
As an initial approach to address \textit{SfC}, we proposed a general-purpose method in this study.
However, in future studies, it would be interesting to develop methods that are specifically tailored to individual materials.

\begin{table}[t]
  \centering
  \footnotesize
  \setlength{\tabcolsep}{4pt}
  \begin{tabularx}{\columnwidth}{lCCCCC}
    \toprule
    \multicolumn{1}{c}{$\hat{E}$} & $2.5 \times 10^5$ & $5.0 \times 10^5$ & $1.0 \times 10^6$ & $2.0 \times 10^6$ & $4.0 \times 10^6$
    \\ \midrule
    Static
    & 0.920
    & 0.921
    & 0.920
    & 0.920
    & 0.920
    \\ \midrule
    \textstrong{SfC-NeRF}
    & \first{0.289}
    & \first{0.254}
    & \first{0.195}
    & \first{0.314}
    & \first{0.374}
    \\ \bottomrule
    \toprule
    \multicolumn{1}{c}{$\hat{\nu}$} & $0.2$ & $0.25$ & $0.3$ & $0.35$ & $0.4$
    \\ \midrule
    Static
    & 0.920
    & 0.919
    & 0.920
    & 0.920
    & 0.921
    \\ \midrule
    \textstrong{SfC-NeRF}
    & \first{0.196}
    & \first{0.198}
    & \first{0.195}
    & \first{0.207}
    & \first{0.224}
    \\ \bottomrule
  \end{tabularx}
  \vspace{-3mm}
  \caption{Comparison of CD ($\times 10^3$$\downarrow$) when varying Young's moduls $\hat{E}$ and Poisson's ratio $\hat{\nu}$.}
  \vspace{-2mm}
  \label{tab:young_poisson}
\end{table}

\begin{table}[t]
  \centering
  \footnotesize
  \setlength{\tabcolsep}{2pt}
  \begin{tabularx}{\columnwidth}{lCCCC}
    \toprule
    & Newtonian & \!\!\!Non-Newtonian\!\!\! & Plasticine & Sand
    \\ \midrule
    Static
    & 0.921 & 0.919 & 0.920 & 0.920
    \\ \midrule
    \textstrong{SfC-NeRF}
    & \first{0.196} & \first{0.218} & \first{0.230} & \first{0.222}
    \\ \bottomrule
  \end{tabularx}
  \vspace{-3mm}
  \caption{Comparison of CD ($\times 10^3$$\downarrow$) for various materials.}
  \vspace{-4mm}
  \label{tab:material}
\end{table}

\subsection{Application to future prediction}
\label{subsec:application}

To demonstrate the practical importance of \textit{SfC}, we investigated the effectiveness of \textit{SfC-NeRF} for future prediction.
Specifically, the first 14 frames were used for training and the subsequent 14 frames were used for evaluation.
We compared \textit{SfC-NeRF}, which \textit{optimizes the internal structures} with \textit{fixed physical properties}, with PAC-NeRF~\cite{XLiICLR2023}, which \textit{optimizes physical properties} with \textit{fixed (filled) internal structures}.
Table~\ref{tab:future} summarizes the results.
SfC-NeRF outperformed PAC-NeRF in terms of the peak-to-signal ratio (PSNR) and structural similarity index measure (SSIM)~\cite{ZWangTIP2004}.
These results indicate that the optimization of the internal structure is crucial in practical scenarios.

\begin{table}[t]
  \centering
  \footnotesize
  \setlength{\tabcolsep}{7.5pt}
  \begin{tabularx}{1\columnwidth}{lcCC}
    \toprule
    & Internal structure & PSNR$\uparrow$ & SSIM$\uparrow$
    \\ \midrule
    PAC-NeRF & Fixed (filled)
    & 23.44 & 0.975
    \\
    \textstrong{SfC-NeRF} & Optimized
    & \first{26.60} & \first{0.981}
    \\ \bottomrule
  \end{tabularx}
  \vspace{-3mm}
  \caption{Results of future prediction.
    The scores were averaged over all cavity sizes and locations for the 40 objects examined in Experiments I and II.}
  \vspace{-1mm}
  \label{tab:future}
\end{table}

\section{Discussion}
\label{sec:discussion}

Based on the above experiments, we obtained promising results for \textit{SfC}.
However, the proposed method has some limitations.
(1) Our approach assumes that the objects deform during collisions.
Therefore, its performance depends on the type of material used.
For example, it may be difficult to apply this method to metallic objects that do not deform.
However, detecting small changes may help to overcome this issue.
(2) Since \textit{SfC} is a novel task, this study focused on evaluating its fundamental performance using simulation data, leaving the validation with real data as a challenge for future research. 
To explore its potential use with real data, we examined its robustness against inaccurate physical properties.
Table~\ref{tab:robustness} presents the results when errors exist in the physical properties.
A significant error (e.g., $-30\%$) in $\hat{\rho}$ causes a notable degradation owing to its negative impact on volume estimation in $\mathcal{L}_{\text{mass}}$.
However, in other cases, the degradation is moderate.
All the scores exceed those of the baselines listed in Table~\ref{tab:size} (e.g., 0.841 by LPO).
These results indicate that the proposed method is robust against inaccurate physical properties.
Additional challenges associated with real data are discussed in Appendix~\ref{subsec:challenge_real}.

\begin{table}[t]
  \centering
  \footnotesize
  \setlength{\tabcolsep}{0.2pt}
  \begin{tabularx}{\columnwidth}{cCCCCCCC}
    \toprule
    Error rate & $-30\%$ & $-20\%$ & $-10\%$ & $0\%$ & $10\%$ & $20\%$ & $30\%$
    \\ \midrule
    Young's modulus $\hat{E}$
    & 0.363 & 0.242 & \third{0.216} & \first{0.195} & \second{0.213} & 0.231 & 0.244
    \\
    Poisson's ratio $\hat{\nu}$
    & 0.240 & 0.231 & \third{0.208} & \first{0.195} & \second{0.200} & 0.214 & 0.236
    \\
    Density $\hat{\rho}$
    & 0.798 & 0.533 & \third{0.289} & \first{0.195} & \second{0.207} & 0.259 & 0.308
    \\ \bottomrule
  \end{tabularx}
  \vspace{-3mm}
  \caption{Comparison of CD ($\times 10^3$$\downarrow$) for inaccurate physical properties.
  In the 0\% case, an elastic material with default settings ($s_c = \left( \frac{2}{3} \right)^3$, $l_c = \text{center}$, $\hat{E} = 10^6$, and $\hat{\nu} = 0.3$) was used.}
  \vspace{-3mm}
  \label{tab:robustness}
\end{table}

\section{Conclusion}
\label{sec:conclusion}

We introduced \textit{SfC} to identify the \textit{invisible internal} structure of an object---a task that remains challenging even with the latest neural 3D representations.
We proposed \textit{SfC-NeRF} as an initial model to address this challenge.
\textit{SfC-NeRF} solves \textit{SfC} by optimizing the internal structures under \textit{physical}, \textit{appearance-preserving}, and \textit{keyframe constraints}, along with \textit{volume annealing}.
As discussed in Section~\ref{sec:discussion}, the proposed method has certain limitations.
Nonetheless, this study suggests a new direction for the development of neural 3D representations, and we believe that future developments in this field will overcome these limitations.

\clearpage

\renewcommand{\baselinestretch}{1}\selectfont

{
    \small
    \bibliographystyle{ieeenat_fullname}
    \bibliography{refs}

\begin{thebibliography}{81}
\providecommand{\natexlab}[1]{#1}
\providecommand{\url}[1]{\texttt{#1}}
\expandafter\ifx\csname urlstyle\endcsname\relax
  \providecommand{\doi}[1]{doi: #1}\else
  \providecommand{\doi}{doi: \begingroup \urlstyle{rm}\Url}\fi

\bibitem[Abou-Chakra et~al.(2024{\natexlab{a}})Abou-Chakra, Dayoub, and
  S{\"u}nderhauf]{JAbouWACV2024}
Jad Abou-Chakra, Feras Dayoub, and Niko S{\"u}nderhauf.
\newblock {ParticleNeRF}: A particle-based encoding for online neural radiance
  fields.
\newblock In \emph{WACV}, 2024{\natexlab{a}}.

\bibitem[Abou-Chakra et~al.(2024{\natexlab{b}})Abou-Chakra, Rana, Dayoub, and
  S{\"u}nderhauf]{JAbouCoRL2024}
Jad Abou-Chakra, Krishan Rana, Feras Dayoub, and Niko S{\"u}nderhauf.
\newblock Physically embodied {Gaussian} splatting: Embedding physical priors
  into a visual {3D} world model for robotics.
\newblock In \emph{CoRL}, 2024{\natexlab{b}}.

\bibitem[Attal et~al.(2022)Attal, Huang, Zollhoefer, Kopf, and
  Kim]{BAttalCVPR2022}
Benjamin Attal, Jia-Bin Huang, Michael Zollhoefer, Johannes Kopf, and Changil
  Kim.
\newblock Learning neural light fields with ray-space embedding networks.
\newblock In \emph{CVPR}, 2022.

\bibitem[Barron et~al.(2021)Barron, Mildenhall, Tancik, Hedman, Martin-Brualla,
  and Srinivasan]{JBarronICCV2021}
Jonathan~T. Barron, Ben Mildenhall, Matthew Tancik, Peter Hedman, Ricardo
  Martin-Brualla, and Pratul~P. Srinivasan.
\newblock {Mip-NeRF}: A multiscale representation for anti-aliasing neural
  radiance fields.
\newblock In \emph{ICCV}, 2021.

\bibitem[Barron et~al.(2022)Barron, Mildenhall, Verbin, Srinivasan, and
  Hedman]{JBarronCVPR2022}
Jonathan~T. Barron, Ben Mildenhall, Dor Verbin, Pratul~P. Srinivasan, and Peter
  Hedman.
\newblock {Mip-NeRF 360}: Unbounded anti-aliased neural radiance fields.
\newblock In \emph{CVPR}, 2022.

\bibitem[Barron et~al.(2023)Barron, Mildenhall, Verbin, Srinivasan, and
  Hedman]{TBarronICCV2023}
Jonathan~T. Barron, Ben Mildenhall, Dor Verbin, Pratul~P. Srinivasan, and Peter
  Hedman.
\newblock {Zip-NeRF}: Anti-aliased grid-based neural radiance fields.
\newblock In \emph{ICCV}, 2023.

\bibitem[Chan et~al.(2021)Chan, Monteiro, Kellnhofer, Wu, and
  Wetzstein]{EChanCVPR2021}
Eric~R. Chan, Marco Monteiro, Petr Kellnhofer, Jiajun Wu, and Gordon Wetzstein.
\newblock {pi-GAN}: Periodic implicit generative adversarial networks for
  {3D}-aware image synthesis.
\newblock In \emph{CVPR}, 2021.

\bibitem[Chan et~al.(2022)Chan, Lin, Chan, Nagano, Pan, De~Mello, Gallo,
  Guibas, Tremblay, Khamis, Karras, and Wetzstein]{EChanCVPR2022}
Eric~R. Chan, Connor~Z. Lin, Matthew~A. Chan, Koki Nagano, Boxiao Pan, Shalini
  De~Mello, Orazio Gallo, Leonidas~J. Guibas, Jonathan Tremblay, Sameh Khamis,
  Tero Karras, and Gordon Wetzstein.
\newblock Efficient geometry-aware {3D} generative adversarial networks.
\newblock In \emph{CVPR}, 2022.

\bibitem[Chan et~al.(2023)Chan, Nagano, Chan, Bergman, Park, Levy, Aittala,
  De~Mello, Karras, and Wetzstein]{ERChanICCV2023}
Eric~R. Chan, Koki Nagano, Matthew~A. Chan, Alexander~W. Bergman, Jeong~Joon
  Park, Axel Levy, Miika Aittala, Shalini De~Mello, Tero Karras, and Gordon
  Wetzstein.
\newblock Generative novel view synthesis with {3D}-aware diffusion models.
\newblock In \emph{ICCV}, 2023.

\bibitem[Chen et~al.(2022)Chen, Xu, Geiger, Yu, and Su]{AChenECCV2022}
Anpei Chen, Zexiang Xu, Andreas Geiger, Jingyi Yu, and Hao Su.
\newblock {TensoRF}: Tensorial radiance fields.
\newblock In \emph{ECCV}, 2022.

\bibitem[Chen et~al.(2023)Chen, Chen, Jiao, and Jia]{RChenICCV2023}
Rui Chen, Yongwei Chen, Ningxin Jiao, and Kui Jia.
\newblock {Fantasia3D}: Disentangling geometry and appearance for high-quality
  text-to-{3D} content creation.
\newblock In \emph{ICCV}, 2023.

\bibitem[Chen et~al.(2024)Chen, Wu, Lin, Harandi, and Cai]{YChenECCV2024}
Yihang Chen, Qianyi Wu, Weiyao Lin, Mehrtash Harandi, and Jianfei Cai.
\newblock {HAC}: Hash-grid assisted context for {3D Gaussian} splatting
  compression.
\newblock In \emph{ECCV}, 2024.

\bibitem[Chu et~al.(2022)Chu, Liu, Zheng, Franz, Seidel, Theobalt, and
  Zayer]{MChuTOG2022}
Mengyu Chu, Lingjie Liu, Quan Zheng, Erik Franz, Hans-Peter Seidel, Christian
  Theobalt, and Rhaleb Zayer.
\newblock Physics informed neural fields for smoke reconstruction with sparse
  data.
\newblock \emph{ACM Trans. Graph.}, 41\penalty0 (4), 2022.

\bibitem[Deng et~al.(2022)Deng, Yang, Xiang, and Tong]{YDengCVPR2022}
Yu Deng, Jiaolong Yang, Jianfeng Xiang, and Xin Tong.
\newblock {GRAM}: Generative radiance manifolds for {3D}-aware image
  generation.
\newblock In \emph{CVPR}, 2022.

\bibitem[Feng et~al.(2023)Feng, Shang, Li, Shao, Jiang, and
  Yang]{YFengCVPR2024}
Yutao Feng, Yintong Shang, Xuan Li, Tianjia Shao, Chenfanfu Jiang, and Yin
  Yang.
\newblock {PIE-NeRF}: Physics-based interactive elastodynamics with {NeRF}.
\newblock In \emph{CVPR}, 2023.

\bibitem[Fridovich-Keil et~al.(2022)Fridovich-Keil, Yu, Tancik, Chen, Recht,
  and Kanazawa]{SFridovichCVPR2022}
Sara Fridovich-Keil, Alex Yu, Matthew Tancik, Qinhong Chen, Benjamin Recht, and
  Angjoo Kanazawa.
\newblock Plenoxels: Radiance fields without neural networks.
\newblock In \emph{CVPR}, 2022.

\bibitem[Gafni et~al.(2021)Gafni, Thies, Zollhofer, and
  Nie{\ss}ner]{GGafniCVPR2021}
Guy Gafni, Justus Thies, Michael Zollhofer, and Matthias Nie{\ss}ner.
\newblock Dynamic neural radiance fields for monocular {4D} facial avatar
  reconstruction.
\newblock In \emph{CVPR}, 2021.

\bibitem[Gao et~al.(2024)Gao, Holynski, Henzler, Brussee, Martin-Brualla,
  Srinivasan, Barron, and Poole]{RGaoNeurIPS2024}
Ruiqi Gao, Aleksander Holynski, Philipp Henzler, Arthur Brussee, Ricardo
  Martin-Brualla, Pratul Srinivasan, Jonathan~T. Barron, and Ben Poole.
\newblock {CAT3D}: Create anything in {3D} with multi-view diffusion models.
\newblock In \emph{NeurIPS}, 2024.

\bibitem[Garbin et~al.(2021)Garbin, Kowalski, Johnson, Shotton, and
  Valentin]{SGarbinICCV2021}
Stephan~J. Garbin, Marek Kowalski, Matthew Johnson, Jamie Shotton, and Julien
  Valentin.
\newblock {FastNeRF}: High-fidelity neural rendering at {200FPS}.
\newblock In \emph{ICCV}, 2021.

\bibitem[Gu et~al.(2022)Gu, Liu, Wang, and Theobalt]{JGuICLR2022}
Jiatao Gu, Lingjie Liu, Peng Wang, and Christian Theobalt.
\newblock {StyleNeRF}: A style-based {3D}-aware generator for high-resolution
  image synthesis.
\newblock In \emph{ICLR}, 2022.

\bibitem[Guan et~al.(2022)Guan, Deng, Wang, and Yang]{SGuanICML2022}
Shanyan Guan, Huayu Deng, Yunbo Wang, and Xiaokang Yang.
\newblock {NeuroFluid}: Fluid dynamics grounding with particle-driven neural
  radiance fields.
\newblock In \emph{ICML}, 2022.

\bibitem[Hedman et~al.(2021)Hedman, Srinivasan, Mildenhall, Barron, and
  Debevec]{PHedmanICCV2021}
Peter Hedman, Pratul~P. Srinivasan, Ben Mildenhall, Jonathan~T. Barron, and
  Paul Debevec.
\newblock Baking neural radiance fields for real-time view synthesis.
\newblock In \emph{ICCV}, 2021.

\bibitem[Hu et~al.(2022)Hu, Liu, Chen, Shen, and Jia]{THuCVPR2022}
Tao Hu, Shu Liu, Yilun Chen, Tiancheng Shen, and Jiaya Jia.
\newblock {EfficientNeRF}: Efficient neural radiance fields.
\newblock In \emph{CVPR}, 2022.

\bibitem[Hu et~al.(2023)Hu, Wang, Ma, Yang, Gao, Liu, and Ma]{WHuICCV2023}
Wenbo Hu, Yuling Wang, Lin Ma, Bangbang Yang, Lin Gao, Xiao Liu, and Yuewen Ma.
\newblock {Tri-MipRF}: {Tri-Mip} representation for efficient anti-aliasing
  neural radiance fields.
\newblock In \emph{ICCV}, 2023.

\bibitem[Hu et~al.(2018)Hu, Fang, Ge, Qu, Zhu, Pradhana, and Jiang]{YHuTOG2018}
Yuanming Hu, Yu Fang, Ziheng Ge, Ziyin Qu, Yixin Zhu, Andre Pradhana, and
  Chenfanfu Jiang.
\newblock A moving least squares material point method with displacement
  discontinuity and two-way rigid body coupling.
\newblock \emph{ACM Trans. Graph.}, 37\penalty0 (4), 2018.

\bibitem[Hu et~al.(2020)Hu, Anderson, Li, Sun, Carr, Ragan-Kelley, and
  Durand]{YHuICLR2020}
Yuanming Hu, Luke Anderson, Tzu-Mao Li, Qi Sun, Nathan Carr, Jonathan
  Ragan-Kelley, and Fr{\'e}do Durand.
\newblock {DiffTaichi}: Differentiable programming for physical simulation.
\newblock In \emph{ICLR}, 2020.

\bibitem[Jiang et~al.(2024{\natexlab{a}})Jiang, Tu, Liu, Gao, Long, Wang, and
  Ma]{YJiangCVPR2024}
Yingwenqi Jiang, Jiadong Tu, Yuan Liu, Xifeng Gao, Xiaoxiao Long, Wenping Wang,
  and Yuexin Ma.
\newblock {GaussianShader}: {3D Gaussian} splatting with shading functions for
  reflective surfaces.
\newblock In \emph{CVPR}, 2024{\natexlab{a}}.

\bibitem[Jiang et~al.(2024{\natexlab{b}})Jiang, Yu, Xie, Li, Feng, Wang, Li,
  Lau, Gao, Yang, and Jiang]{YJiangTOG2024}
Ying Jiang, Chang Yu, Tianyi Xie, Xuan Li, Yutao Feng, Huamin Wang, Minchen Li,
  Henry Lau, Feng Gao, Yin Yang, and Chenfanfu Jiang.
\newblock {VR-GS}: A physical dynamics-aware interactive {Gaussian} splatting
  system in virtual reality.
\newblock \emph{ACM Trans. Graph.}, 78, 2024{\natexlab{b}}.

\bibitem[Kaneko(2022)]{TKanekoCVPR2022}
Takuhiro Kaneko.
\newblock {AR-NeRF}: Unsupervised learning of depth and defocus effects from
  natural images with aperture rendering neural radiance fields.
\newblock In \emph{CVPR}, 2022.

\bibitem[Kaneko(2023)]{TKanekoICCV2023}
Takuhiro Kaneko.
\newblock {MIMO-NeRF}: Fast neural rendering with multi-input multi-output
  neural radiance fields.
\newblock In \emph{ICCV}, 2023.

\bibitem[Kaneko(2024)]{TKanekoCVPR2024}
Takuhiro Kaneko.
\newblock Improving physics-augmented continuum neural radiance field-based
  geometry-agnostic system identification with {Lagrangian} particle
  optimization.
\newblock In \emph{CVPR}, 2024.

\bibitem[Kerbl et~al.(2023)Kerbl, Kopanas, Leimk{\"u}hler, and
  Drettakis]{BKerblTOG2023}
Bernhard Kerbl, Georgios Kopanas, Thomas Leimk{\"u}hler, and George Drettakis.
\newblock {3D Gaussian} splatting for real-time radiance field rendering.
\newblock \emph{ACM Trans. Graph.}, 42\penalty0 (4), 2023.

\bibitem[Kingma and Ba(2015)]{DPKingmaICLR2015}
Diederik~P. Kingma and Jimmy Ba.
\newblock Adam: A method for stochastic optimization.
\newblock In \emph{ICLR}, 2015.

\bibitem[Kurz et~al.(2022)Kurz, Neff, Lv, Zollh{\"o}fer, and
  Steinberger]{AKurzECCV2022}
Andreas Kurz, Thomas Neff, Zhaoyang Lv, Michael Zollh{\"o}fer, and Markus
  Steinberger.
\newblock {AdaNeRF}: Adaptive sampling for real-time rendering of neural
  radiance fields.
\newblock In \emph{ECCV}, 2022.

\bibitem[Lee et~al.(2024)Lee, Rho, Sun, Ko, and Park]{JLeeCVPR2024}
Joo~Chan Lee, Daniel Rho, Xiangyu Sun, Jong~Hwan Ko, and Eunbyung Park.
\newblock Compact {3D Gaussian} representation for radiance field.
\newblock In \emph{CVPR}, 2024.

\bibitem[Li et~al.(2023)Li, Qiao, Chen, Jatavallabhula, Lin, Jiang, and
  Gan]{XLiICLR2023}
Xuan Li, Yi-Ling Qiao, Peter~Yichen Chen, Krishna~Murthy Jatavallabhula, Ming
  Lin, Chenfanfu Jiang, and Chuang Gan.
\newblock {PAC-NeRF}: Physics augmented continuum neural radiance fields for
  geometry-agnostic system identification.
\newblock In \emph{ICLR}, 2023.

\bibitem[Li et~al.(2024)Li, Lyu, Di, Zhai, Lee, and Tombari]{YLiECCV2024}
Yanyan Li, Chenyu Lyu, Yan Di, Guangyao Zhai, Gim~Hee Lee, and Federico
  Tombari.
\newblock {GeoGaussian}: Geometry-aware {Gaussian} splatting for scene
  rendering.
\newblock In \emph{ECCV}, 2024.

\bibitem[Li et~al.(2021)Li, Niklaus, Snavely, and Wang]{ZLiCVPR2021}
Zhengqi Li, Simon Niklaus, Noah Snavely, and Oliver Wang.
\newblock Neural scene flow fields for space-time view synthesis of dynamic
  scenes.
\newblock In \emph{CVPR}, 2021.

\bibitem[Liang et~al.(2024)Liang, Zhang, Hu, Feng, Zhu, and
  Jia]{ZLiangECCV2024}
Zhihao Liang, Qi Zhang, Wenbo Hu, Ying Feng, Lei Zhu, and Kui Jia.
\newblock {Analytic-Splatting}: Anti-aliased {3D Gaussian} splatting via
  analytic integration.
\newblock In \emph{ECCV}, 2024.

\bibitem[Lin et~al.(2023)Lin, Gao, Tang, Takikawa, Zeng, Huang, Kreis, Fidler,
  Liu, and Lin]{CHLinCVPR2023}
Chen-Hsuan Lin, Jun Gao, Luming Tang, Towaki Takikawa, Xiaohui Zeng, Xun Huang,
  Karsten Kreis, Sanja Fidler, Ming-Yu Liu, and Tsung-Yi Lin.
\newblock {Magic3D}: High-resolution text-to-{3D} content creation.
\newblock In \emph{CVPR}, 2023.

\bibitem[Lin et~al.(2021)Lin, Ryabtsev, Sengupta, Curless, Seitz, and
  Kemelmacher-Shlizerman]{SLinCVPR2021}
Shanchuan Lin, Andrey Ryabtsev, Soumyadip Sengupta, Brian~L. Curless, Steven~M.
  Seitz, and Ira Kemelmacher-Shlizerman.
\newblock Real-time high-resolution background matting.
\newblock In \emph{CVPR}, 2021.

\bibitem[Lindell et~al.(2021)Lindell, Martel, and Wetzstein]{DLindellCVPR2021}
David~B. Lindell, Julien N.~P. Martel, and Gordon Wetzstein.
\newblock {AutoInt}: Automatic integration for fast neural volume rendering.
\newblock In \emph{CVPR}, 2021.

\bibitem[Liu et~al.(2024)Liu, Tang, Cheng, Yang, Li, Liu, Huang, Lin, Liu, Wu,
  Xu, and Yuan]{JLiuECCV2024}
Jiayue Liu, Xiao Tang, Freeman Cheng, Roy Yang, Zhihao Li, Jianzhuang Liu, Yi
  Huang, Jiaqi Lin, Shiyong Liu, Xiaofei Wu, Songcen Xu, and Chun Yuan.
\newblock {MirrorGaussian}: Reflecting {3D Gaussians} for reconstructing mirror
  reflections.
\newblock In \emph{ECCV}, 2024.

\bibitem[Liu et~al.(2020)Liu, Gu, Zaw~Lin, Chua, and Theobalt]{LLiuNeurIPS2020}
Lingjie Liu, Jiatao Gu, Kyaw Zaw~Lin, Tat-Seng Chua, and Christian Theobalt.
\newblock Neural sparse voxel fields.
\newblock In \emph{NeurIPS}, 2020.

\bibitem[Lu et~al.(2024)Lu, Guo, Hui, Chen, Yang, Tang, Zhu, and
  Dai]{ZLuCVPR2024}
Zhicheng Lu, Xiang Guo, Le Hui, Tianrui Chen, Min Yang, Xiao Tang, Feng Zhu,
  and Yuchao Dai.
\newblock {3D} geometry-aware deformable {Gaussian} splatting for dynamic view
  synthesis.
\newblock In \emph{CVPR}, 2024.

\bibitem[Luiten et~al.(2024)Luiten, Kopanas, Leibe, and
  Ramanan]{JLuiten3DV2024}
Jonathon Luiten, Georgios Kopanas, Bastian Leibe, and Deva Ramanan.
\newblock Dynamic {3D Gaussians}: Tracking by persistent dynamic view
  synthesis.
\newblock In \emph{3DV}, 2024.

\bibitem[Mildenhall et~al.(2020)Mildenhall, Srinivasan, Tancik, Barron,
  Ramamoorthi, and Ng]{BMildenhallECCV2020}
Ben Mildenhall, Pratul~P. Srinivasan, Matthew Tancik, Jonathan~T. Barron, Ravi
  Ramamoorthi, and Ren Ng.
\newblock {NeRF}: Representing scenes as neural radiance fields for view
  synthesis.
\newblock In \emph{ECCV}, 2020.

\bibitem[Mildenhall et~al.(2022)Mildenhall, Hedman, Martin-Brualla, Srinivasan,
  and Barron]{BMildenhallCVPR2022}
Ben Mildenhall, Peter Hedman, Ricardo Martin-Brualla, Pratul~P. Srinivasan, and
  Jonathan~T. Barron.
\newblock {NeRF} in the dark: High dynamic range view synthesis from noisy raw
  images.
\newblock In \emph{CVPR}, 2022.

\bibitem[M{\"u}ller et~al.(2022)M{\"u}ller, Evans, Schied, and
  Keller]{TMullerTOG2022}
Thomas M{\"u}ller, Alex Evans, Christoph Schied, and Alexander Keller.
\newblock Instant neural graphics primitives with a multiresolution hash
  encoding.
\newblock \emph{ACM Trans. Graph.}, 41\penalty0 (4), 2022.

\bibitem[Neff et~al.(2021)Neff, Stadlbauer, Parger, Kurz, Mueller, Chaitanya,
  Kaplanyan, and Steinberger]{TNeffCVF2021}
Thomas Neff, Pascal Stadlbauer, Mathias Parger, Andreas Kurz, Joerg~H. Mueller,
  Chakravarty R.~Alla Chaitanya, Anton Kaplanyan, and Markus Steinberger.
\newblock {DONeRF}: Towards real-time rendering of compact neural radiance
  fields using depth oracle networks.
\newblock \emph{Comput. Graph. Forum}, 40\penalty0 (4), 2021.

\bibitem[Niedermayr et~al.(2024)Niedermayr, Stumpfegger, and
  Westermann]{SNiedermayrCVPR2024}
Simon Niedermayr, Josef Stumpfegger, and R{\"u}diger Westermann.
\newblock Compressed {3D Gaussian} splatting for accelerated novel view
  synthesis.
\newblock In \emph{CVPR}, 2024.

\bibitem[Niemeyer and Geiger(2021)]{MNiemeyerCVPR2021}
Michael Niemeyer and Andreas Geiger.
\newblock {GIRAFFE}: Representing scenes as compositional generative neural
  feature fields.
\newblock In \emph{CVPR}, 2021.

\bibitem[Park et~al.(2021)Park, Sinha, Barron, Bouaziz, Goldman, Seitz, and
  Martin-Brualla]{KParkICCV2021}
Keunhong Park, Utkarsh Sinha, Jonathan~T. Barron, Sofien Bouaziz, Dan~B.
  Goldman, Steven~M. Seitz, and Ricardo Martin-Brualla.
\newblock Nerfies: Deformable neural radiance fields.
\newblock In \emph{ICCV}, 2021.

\bibitem[Poole et~al.(2023)Poole, Jain, Barron, and Mildenhall]{BPooleICLR2023}
Ben Poole, Ajay Jain, Jonathan~T. Barron, and Ben Mildenhall.
\newblock {DreamFusion}: Text-to-{3D} using {2D} diffusion.
\newblock In \emph{ICLR}, 2023.

\bibitem[Pumarola et~al.(2021)Pumarola, Corona, Pons-Moll, and
  Moreno-Noguer]{APumarolaCVPR2021}
Albert Pumarola, Enric Corona, Gerard Pons-Moll, and Francesc Moreno-Noguer.
\newblock {D-NeRF}: Neural radiance fields for dynamic scenes.
\newblock In \emph{CVPR}, 2021.

\bibitem[Qiu et~al.(2024)Qiu, Yang, Zeng, and Wang]{RZQiuECCV2024}
Ri-Zhao Qiu, Ge Yang, Weijia Zeng, and Xiaolong Wang.
\newblock {Feature Splatting}: Language-driven physics-based scene synthesis
  and editing.
\newblock In \emph{ECCV}, 2024.

\bibitem[Rebain et~al.(2021)Rebain, Jiang, Yazdani, Li, Yi, and
  Tagliasacchi]{DRebainCVPR2021}
Daniel Rebain, Wei Jiang, Soroosh Yazdani, Ke Li, Kwang~Moo Yi, and Andrea
  Tagliasacchi.
\newblock {DeRF}: Decomposed radiance fields.
\newblock In \emph{CVPR}, 2021.

\bibitem[Reiser et~al.(2021)Reiser, Peng, Liao, and Geiger]{CReiserICCV2021}
Christian Reiser, Songyou Peng, Yiyi Liao, and Andreas Geiger.
\newblock {KiloNeRF}: Speeding up neural radiance fields with thousands of tiny
  {MLPs}.
\newblock In \emph{ICCV}, 2021.

\bibitem[Schwarz et~al.(2020)Schwarz, Liao, Niemeyer, and
  Geiger]{KSchwarzNeurIPS2020}
Katja Schwarz, Yiyi Liao, Michael Niemeyer, and Andreas Geiger.
\newblock {GRAF}: Generative radiance fields for {3D}-aware image synthesis.
\newblock In \emph{NeurIPS}, 2020.

\bibitem[Sitzmann et~al.(2021)Sitzmann, Rezchikov, Freeman, Tenenbaum, and
  Durand]{VSitzmannNeurIPS2021}
Vincent Sitzmann, Semon Rezchikov, Bill Freeman, Josh Tenenbaum, and Fredo
  Durand.
\newblock Light field networks: Neural scene representations with
  single-evaluation rendering.
\newblock In \emph{NeurIPS}, 2021.

\bibitem[Skorokhodov et~al.(2022)Skorokhodov, Tulyakov, Wang, and
  Wonka]{ISkorokhodovNeurIPS2022}
Ivan Skorokhodov, Sergey Tulyakov, Yiqun Wang, and Peter Wonka.
\newblock {EpiGRAF}: Rethinking training of {3D} {GANs}.
\newblock In \emph{NeurIPS}, 2022.

\bibitem[Suhail et~al.(2022)Suhail, Esteves, Sigal, and
  Makadia]{MSuhailCVPR2022}
Mohammed Suhail, Carlos Esteves, Leonid Sigal, and Ameesh Makadia.
\newblock Light field neural rendering.
\newblock In \emph{CVPR}, 2022.

\bibitem[Sun et~al.(2022)Sun, Sun, and Chen]{CSunCVPR2022}
Cheng Sun, Min Sun, and Hwann-Tzong Chen.
\newblock Direct voxel grid optimization: Super-fast convergence for radiance
  fields reconstruction.
\newblock In \emph{CVPR}, 2022.

\bibitem[Tang et~al.(2024{\natexlab{a}})Tang, Chen, Chen, Wang, Zeng, and
  Liu]{JTangECCV2024}
Jiaxiang Tang, Zhaoxi Chen, Xiaokang Chen, Tengfei Wang, Gang Zeng, and Ziwei
  Liu.
\newblock {LGM}: Large multi-view {Gaussian} model for high-resolution {3D}
  content creation.
\newblock In \emph{ECCV}, 2024{\natexlab{a}}.

\bibitem[Tang et~al.(2024{\natexlab{b}})Tang, Ren, Zhou, Liu, and
  Zeng]{JTangICLR2024}
Jiaxiang Tang, Jiawei Ren, Hang Zhou, Ziwei Liu, and Gang Zeng.
\newblock {DreamGaussian}: Generative {Gaussian} splatting for efficient {3D}
  content creation.
\newblock In \emph{ICLR}, 2024{\natexlab{b}}.

\bibitem[Tretschk et~al.(2021)Tretschk, Tewari, Golyanik, Zollh{\"o}fer,
  Lassner, and Theobalt]{ETretschkICCV2021}
Edgar Tretschk, Ayush Tewari, Vladislav Golyanik, Michael Zollh{\"o}fer,
  Christoph Lassner, and Christian Theobalt.
\newblock Non-rigid neural radiance fields: Reconstruction and novel view
  synthesis of a dynamic scene from monocular video.
\newblock In \emph{ICCV}, 2021.

\bibitem[Verbin et~al.(2022)Verbin, Hedman, Mildenhall, Zickler, Barron, and
  Srinivasan]{DVerbinCVPR2022}
Dor Verbin, Peter Hedman, Ben Mildenhall, Todd Zickler, Jonathan~T. Barron, and
  Pratul~P. Srinivasan.
\newblock {Ref-NeRF}: Structured view-dependent appearance for neural radiance
  fields.
\newblock In \emph{CVPR}, 2022.

\bibitem[Wang et~al.(2022)Wang, Ren, Huang, Olszewski, Chai, Fu, and
  Tulyakov]{HWangECCV2022}
Huan Wang, Jian Ren, Zeng Huang, Kyle Olszewski, Menglei Chai, Yun Fu, and
  Sergey Tulyakov.
\newblock {R2L}: Distilling neural radiance field to neural light field for
  efficient novel view synthesis.
\newblock In \emph{ECCV}, 2022.

\bibitem[Wang et~al.(2004)Wang, Bovik, Sheikh, and Simoncelli]{ZWangTIP2004}
Zhou Wang, Alan~C. Bovik, Hamid~R. Sheikh, and Eero~P. Simoncelli.
\newblock Image quality assessment: From error visibility to structural
  similarity.
\newblock \emph{IEEE Trans. Image Process.}, 13\penalty0 (4), 2004.

\bibitem[Wang et~al.(2023)Wang, Lu, Wang, Bao, Li, Su, and
  Zhu]{ZWangNeurIPS2023}
Zhengyi Wang, Cheng Lu, Yikai Wang, Fan Bao, Chongxuan Li, Hang Su, and Jun
  Zhu.
\newblock {ProlificDreamer}: High-fidelity and diverse text-to-{3D} generation
  with variational score distillation.
\newblock In \emph{NeurIPS}, 2023.

\bibitem[Wizadwongsa et~al.(2021)Wizadwongsa, Phongthawee, Yenphraphai, and
  Suwajanakorn]{SWizadwongsaCVPR2021}
Suttisak Wizadwongsa, Pakkapon Phongthawee, Jiraphon Yenphraphai, and Supasorn
  Suwajanakorn.
\newblock {NeX}: Real-time view synthesis with neural basis expansion.
\newblock In \emph{CVPR}, 2021.

\bibitem[Xie et~al.(2024)Xie, Zong, Qiu, Li, Feng, Yang, and
  Jiang]{TXieCVPR2024}
Tianyi Xie, Zeshun Zong, Yuxing Qiu, Xuan Li, Yutao Feng, Yin Yang, and
  Chenfanfu Jiang.
\newblock {PhysGaussian}: Physics-integrated {3D Gaussians} for generative
  dynamics.
\newblock In \emph{CVPR}, 2024.

\bibitem[Xue et~al.(2022)Xue, Li, Singh, and Lee]{YXueCVPR2022}
Yang Xue, Yuheng Li, Krishna~Kumar Singh, and Yong~Jae Lee.
\newblock {GIRAFFE HD}: A high-resolution {3D}-aware generative model.
\newblock In \emph{CVPR}, 2022.

\bibitem[Yan et~al.(2024)Yan, Low, Chen, and Lee]{ZYanCVPR2024}
Zhiwen Yan, Weng~Fei Low, Yu Chen, and Gim~Hee Lee.
\newblock Multi-scale {3D Gaussian} splatting for anti-aliased rendering.
\newblock In \emph{CVPR}, 2024.

\bibitem[Yang et~al.(2024{\natexlab{a}})Yang, Gao, Zhou, Jiao, Zhang, and
  Jin]{ZYangCVPR2024}
Ziyi Yang, Xinyu Gao, Wen Zhou, Shaohui Jiao, Yuqing Zhang, and Xiaogang Jin.
\newblock Deformable {3D Gaussians} for high-fidelity monocular dynamic scene
  reconstruction.
\newblock In \emph{CVPR}, 2024{\natexlab{a}}.

\bibitem[Yang et~al.(2024{\natexlab{b}})Yang, Yang, Pan, and
  Zhang]{ZYangICLR2024}
Zeyu Yang, Hongye Yang, Zijie Pan, and Li Zhang.
\newblock Real-time photorealistic dynamic scene representation and rendering
  with {4D Gaussian} splatting.
\newblock In \emph{ICLR}, 2024{\natexlab{b}}.

\bibitem[Yi et~al.(2024)Yi, Fang, Wang, Wu, Xie, Zhang, Liu, Tian, and
  Wang]{TYiCVPR2024}
Taoran Yi, Jiemin Fang, Junjie Wang, Guanjun Wu, Lingxi Xie, Xiaopeng Zhang,
  Wenyu Liu, Qi Tian, and Xinggang Wang.
\newblock {GaussianDreamer}: Fast generation from text to {3D Gaussians} by
  bridging {2D} and {3D} diffusion models.
\newblock In \emph{CVPR}, 2024.

\bibitem[Yu et~al.(2021)Yu, Li, Tancik, Li, Ng, and Kanazawa]{AYu2021ICCV}
Alex Yu, Ruilong Li, Matthew Tancik, Hao Li, Ren Ng, and Angjoo Kanazawa.
\newblock {PlenOctrees} for real-time rendering of neural radiance fields.
\newblock In \emph{ICCV}, 2021.

\bibitem[Yu et~al.(2024)Yu, Chen, Huang, Sattler, and Geiger]{ZYuCVPR2024}
Zehao Yu, Anpei Chen, Binbin Huang, Torsten Sattler, and Andreas Geiger.
\newblock {Mip-Splatting}: Alias-free {3D Gaussian} splatting.
\newblock In \emph{CVPR}, 2024.

\bibitem[Zhang et~al.(2020)Zhang, Riegler, Snavely, and
  Koltun]{KZhangArXiv2020}
Kai Zhang, Gernot Riegler, Noah Snavely, and Vladlen Koltun.
\newblock {NeRF++}: Analyzing and improving neural radiance fields.
\newblock \emph{arXiv preprint arXiv:2010.07492}, 2020.

\bibitem[Zhou et~al.(2024)Zhou, Fan, Xu, Chang, Chari, Bharadwaj, You, Wang,
  and Kadambi]{SZhouECCV2024}
Shijie Zhou, Zhiwen Fan, Dejia Xu, Haoran Chang, Pradyumna Chari, Tejas
  Bharadwaj, Suya You, Zhangyang Wang, and Achuta Kadambi.
\newblock {DreamScene360}: Unconstrained text-to-{3D} scene generation with
  panoramic {Gaussian} splatting.
\newblock In \emph{ECCV}, 2024.

\end{thebibliography}
}

\clearpage

\appendix

\begingroup
\hypersetup{linkcolor=black}
\tableofcontents
\endgroup

\section{Detailed analyses and discussions}
\label{sec:detailed_analyses_discussions}

\subsection{Detailed ablation studies}
\label{sec:detailed_ablation}

Owing to space limitations in the main text, we conducted an ablation study that focused only on the selected key components.
In this appendix, we present detailed ablation studies that further assess the effectiveness of the proposed method from multiple perspectives.
Specifically, we examine the effects of \textit{each appearance-preserving loss} (Appendix~\ref{subsubsec:effect_appearance_preserving}), \textit{keyframe selection} (Appendix~\ref{subsubsec:effect_key_frame}), and \textit{background loss} (Appendix~\ref{subsubsec:effect_background_loss}).

\subsubsection{Effect of each appearance-preserving loss}
\label{subsubsec:effect_appearance_preserving}

As explained in Section~\ref{subsec:proposal}, regarding appearance-preserving constraints, we adopted two appearance-preserving losses (APLs): pixel-preserving loss $\mathcal{L}_{\text{pixel}_0}$ (Equation~\ref{eq:pixel_pres_loss}) and depth-preserving loss $\mathcal{L}_{\text{depth}_0}$ (Equation~\ref{eq:depth_pres_loss}).
These losses help prevent the degradation of the external structure, which is effectively learned from the first frame of the video sequence during the fitting process across the entire video sequence.
In the ablation study presented in Sections~\ref{subsec:experiment1} and \ref{subsec:experiment2}, we ablated both losses simultaneously to examine the overall effect of APLs.
For a more detailed ablation study, we assessed the performance when each APL was individually ablated.

\begin{table}[t]
  \centering
  \footnotesize
  \begin{tabularx}{\columnwidth}{ccCCCCC}
    \toprule
    $\mathcal{L}_{\text{pixel}_0}$ & $\mathcal{L}_{\text{depth}_0}$ & $0$ & $(\frac{1}{2})^3$ & $(\frac{2}{3})^3$ & $(\frac{3}{4})^3$ & Avg.
    \\ \midrule
    &
    & 0.106
    & 0.423
    & 0.898
    & 1.326
    & 0.688
    \\
    \checkmark
    &
    & \third{0.105}
    & \second{0.142}
    & \third{0.334}
    & \third{0.342}
    & \second{0.231}
    \\
    & \checkmark
    & \first{0.079}
    & \third{0.313}
    & \second{0.314}
    & \second{0.287}
    & \third{0.248}
    \\ \midrule
    \textstrong{\checkmark}
    & \textstrong{\checkmark}
    & \second{0.081}
    & \first{0.122}
    & \first{0.195}
    & \first{0.262}
    & \first{0.165}
    \\ \bottomrule
  \end{tabularx}
  \vspace{-2mm}
  \caption{Results of the detailed ablation study of APLs when the cavity size $s_c$ is varied.
    The score indicates CD ($\times 10^3$$\downarrow$).
    A checkmark \checkmark\, indicates that the corresponding loss was used.}
  \label{tab:apl_size}
\end{table}

\begin{table}[t]
  \centering
  \footnotesize
  \begin{tabularx}{\columnwidth}{ccCCCCC}
    \toprule
    $\mathcal{L}_{\text{pixel}_0}$ & $\mathcal{L}_{\text{depth}_0}$ & left & right & up & down & Avg.
    \\ \midrule    
    &
    & 0.845
    & 0.783
    & 0.805
    & 0.583
    & 0.754
    \\
    \checkmark
    &
    & \first{0.295}
    & \third{0.451}
    & \second{0.325}
    & \second{0.311}
    & \second{0.345}
    \\
    & \checkmark
    & \third{0.362}
    & \second{0.299}
    & \third{0.348}
    & \third{0.389}
    & \third{0.349}
    \\ \midrule
    \textstrong{\checkmark}
    & \textstrong{\checkmark}
    & \second{0.303}
    & \first{0.258}
    & \first{0.274}
    & \first{0.291}
    & \first{0.281}
    \\ \bottomrule
  \end{tabularx}
  \vspace{-2mm}
  \caption{Results of the detailed ablation study of APLs when the cavity location $l_c$ is varied.
    The score indicates CD ($\times 10^3$$\downarrow$).
    A checkmark \checkmark\, indicates that the corresponding loss was used.}
  \vspace{-4mm}
  \label{tab:apl_location}
\end{table}

\smallskip\noindent
\textbf{Results.}
Table~\ref{tab:apl_size} summarizes the results when the cavity size $s_c$ is varied, and Table~\ref{tab:apl_location} summarizes the results when the cavity location $l_c$ is varied.
Our findings are threefold:

\smallskip\noindent
\textit{(1) No APL vs. either $\mathcal{L}_{\text{pixel}_0}$ or $\mathcal{L}_{\text{depth}_0}$.}
Both SfC-NeRF with only $\mathcal{L}_{\text{pixel}_0}$ and SFC-NeRF with only $\mathcal{L}_{\text{depth}_0}$ outperformed SfC-NeRF without APL in all cases.
These results indicate that both $\mathcal{L}_{\text{pixel}_0}$ and $\mathcal{L}_{\text{depth}_0}$ effectively enhance the performance of \textit{SfC}.

\smallskip\noindent
\textit{(2) Full APLs vs. either $\mathcal{L}_{\text{pixel}_0}$ or $\mathcal{L}_{\text{depth}_0}$.}
SfC-NeRF with both $\mathcal{L}_{\text{pixel}_0}$ and $\mathcal{L}_{\text{depth}_0}$ outperformed SfC-NeRF with only $\mathcal{L}_{\text{pixel}_0}$ and SFC-NeRF with only $\mathcal{L}_{\text{depth}_0}$ in most cases.
These results indicate that $\mathcal{L}_{\text{pixel}_0}$ and $\mathcal{L}_{\text{depth}_0}$ contribute to improving the performance of \textit{SfC} from different perspectives and are most effective when used together.

\smallskip\noindent
\textit{(3) $\mathcal{L}_{\text{pixel}_0}$ vs $\mathcal{L}_{\text{depth}_0}$.}
The superiority or inferiority of each loss depends on the cavity settings.
This is related to the learnability of 3D appearance, and further detailed analyses will be an interesting direction for future research.

\subsubsection{Effect of keyframe selection}
\label{subsubsec:effect_key_frame}

\begin{figure}[t]
  \centering
  \includegraphics[width=0.92\linewidth]{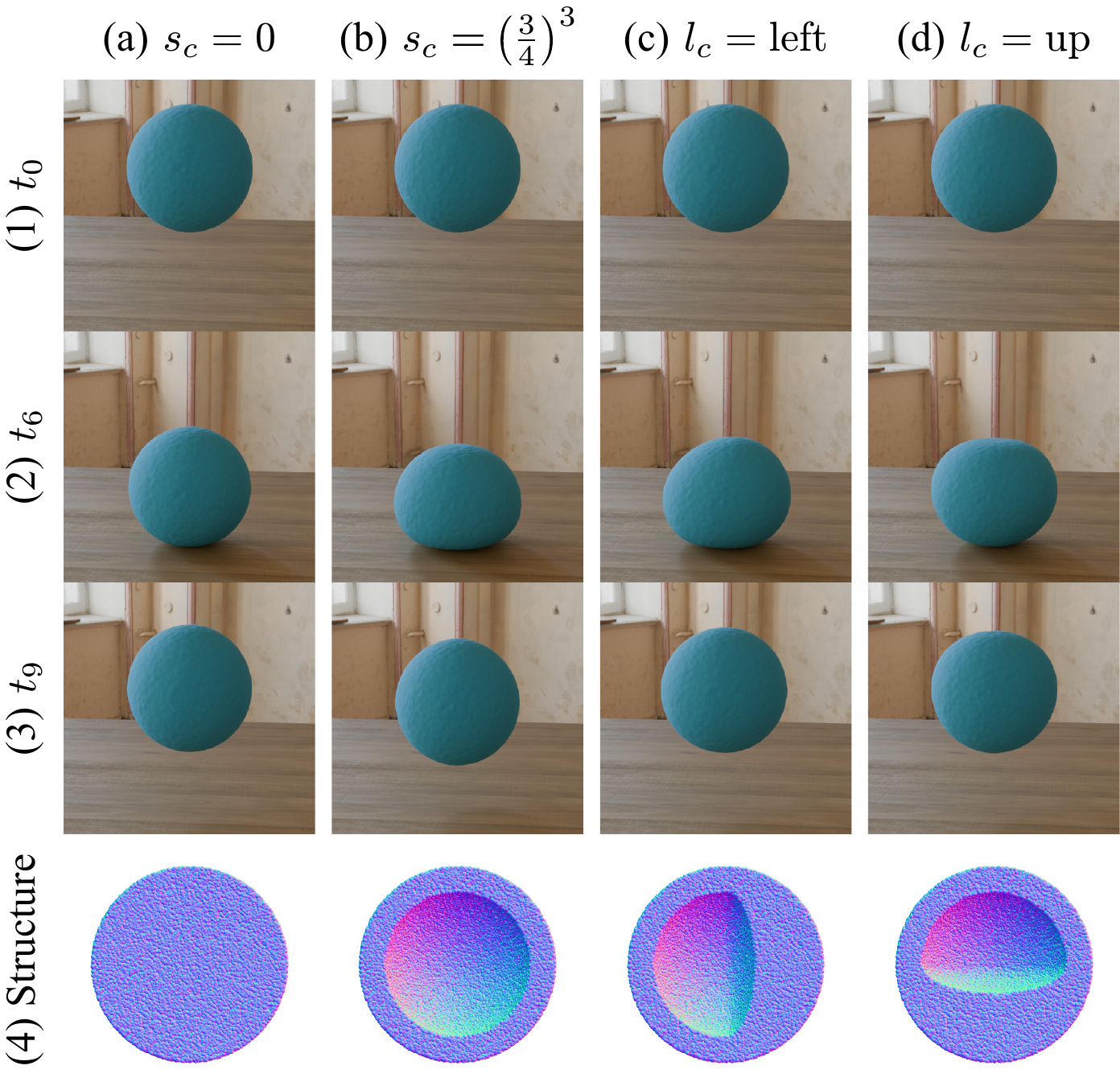}
  \caption{Comparison of appearances for objects with different internal structures when $t$ is varied within $\{ t_0, t_6, t_9 \}$.}
  \vspace{-2mm}
  \label{fig:comparison_appearance}
\end{figure}

As discussed in Section~\ref{subsec:proposal}, regarding keyframe constraints, we employed a keyframe pixel loss $\mathcal{L}_{\text{pixel}_k}$ (Equation~\ref{eq:key_frame_loss}) to effectively capture shape changes caused by internal structures.
Specifically, we selected the frame immediately after the collision as the keyframe ($k = 6$, where $k$ is the keyframe index) for the experiments described in the main text.
An important question is whether the choice of $k$ is optimal.
To investigate this, we evaluated the change in performance by varying the value of $k$, specifically within $\{ 6, 9 \}$.
Figure~\ref{fig:comparison_appearance} compares the appearances of objects with different internal structures in these keyframes.
For reference, we also provide scores for the model without keyframe pixel loss (denoted by \textit{$k = \text{None}$}).

\smallskip\noindent
\textbf{Results.}
Table~\ref{tab:keyframe_size} summarizes the results when the cavity size $s_c$ is varied, and Table~\ref{tab:keyframe_location} summarizes the results when the cavity location $l_c$ is varied.
Our findings are twofold:

\smallskip\noindent
\textit{(1) $\mathcal{L}_{\text{pixel}_6}$ vs. $\mathcal{L}_{\text{pixel}_9}$.}
SfC-NeRF with $\mathcal{L}_{\text{pixel}_6}$ outperformed that with $\mathcal{L}_{\text{pixel}_9}$ in most cases.
As shown in Figure~\ref{fig:comparison_appearance}, immediately after the collision (at $t_6$ (2)), the difference in the shapes of the objects is noticeable.
However, as time progressed after the collision (at $t_9$ (3)), the difference in the shapes of the objects decreased, whereas the difference in their positions became more pronounced.
We consider this to be the main reason why SfC-NeRF with $\mathcal{L}_{\text{pixel}_6}$ performed better than that with $\mathcal{L}_{\text{pixel}_9}$.

\smallskip\noindent
\textit{(2) $\mathcal{L}_{\text{pixel}_6}$/$\mathcal{L}_{\text{pixel}_9}$ vs. None.}
We found that SfC-NeRF with $\mathcal{L}_{\text{pixel}_6}$ or $\mathcal{L}_{\text{pixel}_9}$ outperformed SfC-NeRF without keyframe pixel loss in most cases.
These results indicate that strategically weighing frames is more effective than treating all frames equally.

\begin{table}[t]
  \centering
  \footnotesize
  \begin{tabularx}{\columnwidth}{cCCCCC}
    \toprule
    $k$ & $0$ & $(\frac{1}{2})^3$ & $(\frac{2}{3})^3$ & $(\frac{3}{4})^3$ & Avg.
    \\ \midrule
    None
    & \second{0.082}
    & \third{0.127}
    & \third{0.211}
    & \third{0.325}
    & \third{0.186}
    \\ \midrule
    \textstrong{$6$}
    & \first{0.081}
    & \second{0.122}
    & \first{0.195}
    & \first{0.262}
    & \first{0.165}
    \\
    $9$
    & \second{0.082}
    & \first{0.120}
    & \second{0.208}
    & \second{0.290}
    & \second{0.175}
    \\ \bottomrule
  \end{tabularx}
  \vspace{-2mm}
  \caption{Analysis of the effect of keyframe selection when the cavity size $s_c$ is varied.
    The score indicates CD ($\times 10^3$$\downarrow$).
    When $k = \text{None}$, a keyframe pixel loss $\mathcal{L}_{\text{pixel}_k}$ was not used.
    In contrast, when $k \in \{ 6, 9 \}$, $\mathcal{L}_{\text{pixel}_k}$ was used.}
  \label{tab:keyframe_size}
\end{table}

\begin{table}[t]
  \centering
  \footnotesize
  \begin{tabularx}{\columnwidth}{cCCCCC}
    \toprule
    $k$ & left & right & up & down & Avg.
    \\ \midrule
    None
    & \third{0.308}
    & \second{0.296}
    & \second{0.307}
    & \third{0.313}
    & \third{0.306}
    \\ \midrule
    \textstrong{$6$}
    & \second{0.303}
    & \first{0.258}
    & \first{0.274}
    & \first{0.291}
    & \first{0.281}
    \\
    $9$
    & \first{0.296}
    & \second{0.296}
    & \third{0.313}
    & \second{0.303}
    & \second{0.302}
    \\ \bottomrule
  \end{tabularx}
  \vspace{-2mm}
  \caption{Analysis of the effect of keyframe selection when the cavity location $l_c$ is varied.
    The score indicates CD ($\times 10^3$$\downarrow$).
    When $k = \text{None}$, a keyframe pixel loss $\mathcal{L}_{\text{pixel}_k}$ was not used.
    In contrast, when $k \in \{ 6, 9 \}$, $\mathcal{L}_{\text{pixel}_k}$ was used.}
  \vspace{-4mm}
  \label{tab:keyframe_location}
\end{table}

\subsubsection{Effect of background loss}
\label{subsubsec:effect_background_loss}

As mentioned in the explanation of preprocessing in Section~\ref{subsec:experimental_setup}, we use a background loss $\mathcal{L}_{\text{bg}}$ by leveraging the fact that the background segmentation has been obtained.
For example, when an image with a white background is given, this background loss is useful for distinguishing whether the white part belongs to the background or a foreground object.
We used a background segmentation that is not created manually but is predicted from a given image using a DNN-based image matting model~\cite{SLinCVPR2021}.
Therefore, this setting is not unrealistic.
However, it is important to investigate the effectiveness of the background loss.
To this end, we investigated the performance of \textit{SfC-NeRF$_{-\text{bg}}$}, where the background loss ($\mathcal{L}_{\text{bg}}$) was ablated.
In this setting, the performance of a model trained using only the first frame of the video sequence (Step~(i) in Figure~\ref{fig:pipelines}(a)) also changes because the background loss is ablated in this step.
We refer to this model as \textit{Static$_{-\text{bg}}$}.
We compared the scores of these models with those of the original models (i.e., \textit{SfC-NeRF} and \textit{Static}).

\smallskip\noindent
\textbf{Results.}
Table~\ref{tab:bg_size} summarizes the results when the cavity size $s_c$ is varied, and Table~\ref{tab:bg_location} summarizes the results when the cavity location $l_c$ is varied.
Our findings are twofold:

\smallskip\noindent
\textit{(1) SfC-NeRF vs. SfC-NeRF$_{-\text{bg}}$.}
SfC-NeRF outperformed SfC-NeRF$_{-\text{bg}}$ in most cases.
As mentioned above, the background loss is useful for distinguishing between background and foreground objects, allowing for a more accurate capture of external structures.
The movements of an object are affected by its external and internal structures.
Therefore, if the external structure can be estimated more accurately, the internal structure can also be estimated more accurately.

\smallskip\noindent
\textit{(2) SfC-NeRF$_{-\text{bg}}$ vs Static$_{-\text{bg}}$.}
SfC-NeRF$_{-\text{bg}}$ outperformed Static $_{-\text{bg}}$ except when dealing with filled objects ($s_c = 0$ in Table~\ref{tab:bg_size}).\footnote{When handling a filled object, inaccurate estimation of external structure is problematic because it causes a difference between the actual and estimated masses.
  In this situation, if the estimated mass is encouraged to approach the ground-truth mass using a mass loss while maintaining the external appearance using APLs, the internal structure must be changed unnecessarily.
  Consequently, SfC-NeRF$_{-\text{bg}}$ degrades the performance of \textit{SfC} when handling filled objects.
  An accurate estimation of the external structure using a background loss is effective for addressing this issue.}
These results indicate that the proposed method is effective for improving the performance of \textit{SfC}, even without the use of advanced techniques, such as background loss.

\begin{table}[t]
  \centering
  \footnotesize
  \begin{tabularx}{\columnwidth}{lCCCCC}
    \toprule
    & $0$ & $(\frac{1}{2})^3$ & $(\frac{2}{3})^3$ & $(\frac{3}{4})^3$ & Avg.
    \\ \midrule
    Static
    & \second{0.093}
    & 0.294
    & 0.920
    & 1.574
    & 0.720
    \\
    \textstrong{SfC-NeRF}
    & \first{0.081}
    & \first{0.122}
    & \first{0.195}
    & \first{0.262}
    & \first{0.165}
    \\ \midrule
    Static$_{-\text{bg}}$
    & \second{0.093}
    & 0.290
    & 0.906
    & 1.545
    & 0.708
    \\
    \textstrong{SfC-NeRF$_{-\text{bg}}$}
    & 0.101
    & \second{0.149}
    & \second{0.222}
    & \second{0.279}
    & \second{0.188}
    \\ \bottomrule
  \end{tabularx}
  \vspace{-2mm}
  \caption{Results of the ablation study of background loss when the cavity size $s_c$ is varied.
    The score indicates CD ($\times 10^3$$\downarrow$).}
  \label{tab:bg_size}
\end{table}

\begin{table}[t]
  \centering
  \footnotesize
  \begin{tabularx}{\columnwidth}{lCCCCC}
    \toprule
    & left & right & up & down & Avg.
    \\ \midrule
    Static
    & 0.841
    & 0.842
    & 0.815
    & 0.813
    & 0.828
    \\
    \textstrong{SfC-NeRF}
    & \first{0.303}
    & \second{0.258}
    & \first{0.274}
    & \second{0.291}
    & \first{0.281}
    \\ \midrule
    Static$_{-\text{bg}}$
    & 0.831
    & 0.830
    & 0.799
    & 0.800
    & 0.815
    \\
    \textstrong{SfC-NeRF$_{-\text{bg}}$}
    & \second{0.324}
    & \first{0.210}
    & \second{0.361}
    & \first{0.277}
    & \second{0.293}
    \\ \bottomrule
  \end{tabularx}
  \vspace{-2mm}
  \caption{Results of the ablation study of background loss when the cavity location $l_c$ is varied.
    The score indicates CD ($\times 10^3$$\downarrow$).}
  \vspace{-4mm}
  \label{tab:bg_location}
\end{table}

\subsection{Extended experiments}
\label{subsec:extended_experiments}

\subsubsection{Experiment IV: Influence of collision angle}
\label{subsubsec:experiment4}

In the above experiments, the collision angle was fixed, as shown in Figures~\ref{fig:results_sphere}--\ref{fig:results_material}, regardless of the internal structure and physical properties, to focus on comparisons related to the internal structures and physical properties.
For completeness, we investigated the influence of \textit{collision angle} $\theta_{c}$ on the performance of \textit{SfC}.
Specifically, we selected objects with default settings ($s_c = (\frac{2}{3})^3$, $l_c = \text{center}$, and elastic material defined by $\hat{E} = 1.0 \times 10^6$ and $\hat{\nu} = 0.3$) as the objects of investigation and examined their performance when only the collision angles were altered.
The objects were rotated in the depth direction, as shown in Figure~\ref{fig:results_angle}.
The collision angle $\theta_c$ was chosen from $\{ 0\tcdegree, 22.5\tcdegree, 45\tcdegree, 67.5\tcdegree, 90\tcdegree \}$.
We compared the performance of \textit{Static} and \textit{SfC-NeRF}.

\begin{table}[t]
  \centering
  \footnotesize
  \setlength{\tabcolsep}{4pt}
  \begin{tabularx}{\columnwidth}{lCCCCC}
    \toprule
    Sphere & $0\tcdegree$ & $22.5\tcdegree$ & $45\tcdegree$ & $67.5\tcdegree$ & $90\tcdegree$
    \\ \midrule
    Static
    & 1.164
    & 1.163
    & 1.163
    & 1.162
    & 1.160
    \\ \midrule
    \textstrong{SfC-NeRF}
    & \first{0.067}
    & \first{0.068}
    & \first{0.066}
    & \first{0.067}
    & \first{0.066}
    \\ \bottomrule
    \toprule
    Cube & $0\tcdegree$ & $22.5\tcdegree$ & $45\tcdegree$ & $67.5\tcdegree$ & $90\tcdegree$
    \\ \midrule
    Static
    & 0.775
    & 0.776
    & 0.848
    & 0.768
    & 0.776
    \\ \midrule
    \textstrong{SfC-NeRF}
    & \first{0.201}
    & \first{0.173}
    & \first{0.627}
    & \first{0.201}
    & \first{0.201}
    \\ \bottomrule
    \toprule
    Bicone & $0\tcdegree$ & $22.5\tcdegree$ & $45\tcdegree$ & $67.5\tcdegree$ & $90\tcdegree$
    \\ \midrule
    Static
    & 0.933
    & 0.925
    & 0.918
    & 0.921
    & 0.926
    \\ \midrule
    \textstrong{SfC-NeRF}
    & \first{0.144}
    & \first{0.194}
    & \first{0.187}
    & \first{0.146}
    & \first{0.154}
    \\ \bottomrule
    \toprule
    Cylinder & $0\tcdegree$ & $22.5\tcdegree$ & $45\tcdegree$ & $67.5\tcdegree$ & $90\tcdegree$
    \\ \midrule
    Static
    & 0.891
    & 0.905
    & 0.915
    & 0.905
    & 0.964
    \\ \midrule
    \textstrong{SfC-NeRF}
    & \first{0.342}
    & \first{0.288}
    & \first{0.311}
    & \first{0.209}
    & \first{0.639}
    \\ \bottomrule
    \toprule
    Diamond & $0\tcdegree$ & $22.5\tcdegree$ & $45\tcdegree$ & $67.5\tcdegree$ & $90\tcdegree$
    \\ \midrule
    Static
    & 0.837
    & 0.830
    & 0.833
    & 0.819
    & 0.838
    \\ \midrule
    \textstrong{SfC-NeRF}
    & \first{0.220}
    & \first{0.300}
    & \first{0.222}
    & \first{0.163}
    & \first{0.209}
    \\ \bottomrule
  \end{tabularx}
  \vspace{-2mm}
  \caption{Comparison of CD ($\times 10^3$$\downarrow$) when collision angle $\theta_c$ is varied.}
  \vspace{-4mm}
  \label{tab:angle}
\end{table}

\smallskip\noindent
\textbf{Results.}
Table~\ref{tab:angle} summarizes the quantitative results.
Figure~\ref{fig:results_angle} shows the qualitative results.
Our findings are twofold:

\smallskip\noindent
\textit{(1) SfC-NeRF vs. Static.}
SfC-NeRF outperformed Static in all cases.
These results indicate that optimizing the internal structure through a video sequence using the proposed method is beneficial, regardless of the collision angle.

\smallskip\noindent
\textit{(2) Effect of collision angle.}
We found that the collision angle influenced the performance of \textit{SfC}.
The strength of this effect depends on the object shape.
There are three possible reasons for this performance variation:
\textit{(i) Changes in estimation accuracy of external structures.}
The internal structure was optimized under the constraint that the external structure, learned from the first frame, should be maintained.
Therefore, when the accuracy of the external structure estimation changed, the accuracy of the internal structure estimation also changed.
\textit{(ii) Difference in amount of deformation.}
The amount of deformation varied depending on the collision angle.
This factor also affected the ease of estimating the internal structure.
\textit{(iii) Asymmetry.}
When an object was not symmetrical relative to the collision angle, its behavior after the collision became asymmetrical.
Consequently, the ease of estimating the internal structure also became asymmetrical.

\begin{table*}[t]
  \centering
  \footnotesize
  \begin{tabularx}{\textwidth}{lCCCCCCCCCC}
    \toprule
    & \multicolumn{2}{c}{$0$} & \multicolumn{2}{c}{$(\frac{1}{2})^3$} & \multicolumn{2}{c}{$(\frac{2}{3})^3$} & \multicolumn{2}{c}{$(\frac{3}{4})^3$} & \multicolumn{2}{c}{Avg.}
    \\ \midrule
    Static
    & 0.093 & 0.104
    & 0.294 & 0.309
    & 0.920 & 1.057
    & 1.574 & 1.964
    & 0.720 & 0.859
    \\ \midrule
    GO
    & 0.091 & 0.092
    & 0.301 & 0.301
    & 0.941 & 0.944
    & 1.586 & 1.612
    & 0.730 & 0.737
    \\
    GO$_{\text{mass}}$
    & \first{0.081} & \first{0.083}
    & 0.319 & 0.325
    & 1.244 & 1.266
    & 2.291 & 2.367
    & 0.984 & 1.010
    \\
    LPO
    & 0.092 & 0.091
    & 0.284 & 0.282
    & 0.841 & 0.833
    & 1.406 & 1.380
    & 0.656 & 0.646
    \\
    LPO$_{\text{mass}}$
    & 0.087 & 0.087
    & 0.284 & 0.283
    & 0.876 & 0.868
    & 1.477 & 1.451
    & 0.681 & 0.672
    \\ \midrule
    SfC-NeRF$_{-\text{mass}}$
    & 0.089 & 0.090
    & \third{0.226} & \third{0.225}
    & 0.550 & 0.544
    & 1.148 & 1.112
    & 0.503 & 0.493
    \\
    SfC-NeRF$_{-\text{APL}}$
    & 0.106 & 0.108
    & 0.423 & 0.421
    & 0.898 & 0.886
    & 1.326 & 1.307
    & 0.688 & 0.680
    \\
    SfC-NeRF$_{-\text{APT}}$
    & \third{0.085} & 0.101
    & 0.261 & 0.279
    & \third{0.332} & \third{0.337}
    & 0.661 & 0.680
    & 0.335 & \third{0.349}
    \\
    SfC-NeRF$_{-\text{key}}$
    & \second{0.082} & \third{0.086}
    & \second{0.127} & \second{0.131}
    & \second{0.211} & \second{0.213}
    & \second{0.325} & \second{0.325}
    & \second{0.186} & \second{0.189}
    \\
    SfC-NeRF$_{-\text{VA}}$
    & 0.146 & 0.269
    & 0.293 & 0.338
    & 0.370 & 0.407
    & \third{0.456} & \third{0.485}
    & \third{0.316} & 0.375
    \\ \midrule
    \textstrong{SfC-NeRF}
    & \first{0.081} & \second{0.085}
    & \first{0.122} & \first{0.126}
    & \first{0.195} & \first{0.196}
    & \first{0.262} & \first{0.258}
    & \first{0.165} & \first{0.166}
    \\ \bottomrule
  \end{tabularx}
  \vspace{-2mm}
  \caption{Comparison of CD ($\times 10^3$$\downarrow$) when the cavity size $s_c$ is varied.
    This is an extended version of Table~\ref{tab:size}.
    For each condition, the left score indicates CD$_{\text{static}}$, the chamfer distance between $\mathcal{P}(t_0)$ and $\hat{\mathcal{P}}(t_0)$ at the first frame, i.e., $t = t_0$, and the right score indicates CD$_{\text{video}}$, the chamfer distance between $\mathcal{P}(t)$ and $\hat{\mathcal{P}}(t)$ averaged over the entire video sequence, i.e., $t \in \{ t_0, \dots, t_{N-1} \}$.}
  \label{tab:size_video}
\end{table*}

\begin{table*}[t]
  \centering
  \footnotesize
  \begin{tabularx}{\textwidth}{lCCCCCCCCCC}
    \toprule
    & \multicolumn{2}{c}{left} & \multicolumn{2}{c}{right} & \multicolumn{2}{c}{up} & \multicolumn{2}{c}{down} & \multicolumn{2}{c}{Avg.}
    \\ \midrule
    \multirow{2}{*}{Static}
    & 0.841 & 1.159
    & 0.842 & 1.306
    & 0.815 & 1.731
    & 0.813 & 1.241
    & 0.828 & 1.359
    \\
    & \gray{(0.841)} & \gray{(1.294)}
    & \gray{(0.843)} & \gray{(1.154)}
    & \gray{(0.814)} & \gray{(1.246)}
    & \gray{(0.813)} & \gray{(1.727)}
    & \gray{(0.828)} & \gray{(1.355)}
    \\ \midrule
    \multirow{2}{*}{GO}
    & 0.874 & 0.879
    & 0.853 & 0.870
    & 0.878 & 0.875
    & 0.870 & 1.035
    & 0.869 & 0.915
    \\
    & \gray{(0.872)} & \gray{(2.606)}
    & \gray{(0.856)} & \gray{(2.549)}
    & \gray{(0.881)} & \gray{(1.471)}
    & \gray{(0.870)} & \gray{(1.673)}
    & \gray{(0.870)} & \gray{(2.075)}
    \\
    \multirow{2}{*}{GO$_{\text{mass}}$}
    & 1.349 & 1.386
    & 1.334 & 1.375
    & 1.104 & 1.141
    & 1.001 & 1.370
    & 1.197 & 1.318
    \\
    & \gray{(1.340)} & \gray{(3.134)}
    & \gray{(1.344)} & \gray{(3.126)}
    & \gray{(1.127)} & \gray{(1.866)}
    & \gray{(1.004)} & \gray{(1.805)}
    & \gray{(1.204)} & \gray{(2.483)}
    \\
    \multirow{2}{*}{LPO}
    & 0.791 & 0.789
    & 0.787 & 0.787
    & 0.796 & 0.776
    & 0.743 & 0.721
    & 0.779 & 0.768
    \\
    & \gray{(0.802)} & \gray{(2.493)}
    & \gray{(0.800)} & \gray{(2.507)}
    & \gray{(0.819)} & \gray{(1.468)}
    & \gray{(0.737)} & \gray{(1.471)}
    & \gray{(0.790)} & \gray{(1.985)}
    \\
    \multirow{2}{*}{LPO$_{\text{mass}}$}
    & 0.824 & 0.822
    & 0.817 & 0.818
    & 0.828 & 0.806
    & 0.775 & 0.753
    & 0.811 & 0.800
    \\
    & \gray{(0.833)} & \gray{(2.529)}
    & \gray{(0.832)} & \gray{(2.556)}
    & \gray{(0.847)} & \gray{(1.497)}
    & \gray{(0.771)} & \gray{(1.538)}
    & \gray{(0.821)} & \gray{(2.030)}
    \\ \midrule
    \multirow{2}{*}{SfC-NeRF$_{-\text{mass}}$}
    & \third{0.513} & \third{0.520}
    & 0.485 & \third{0.491}
    & 0.705 & 0.689
    & 0.479 & 0.457
    & 0.545 & \third{0.539}
    \\
    & \gray{(0.858)} & \gray{(2.502)}
    & \gray{(0.878)} & \gray{(2.661)}
    & \gray{(0.747)} & \gray{(1.506)}
    & \gray{(0.956)} & \gray{(1.762)}
    & \gray{(0.860)} & \gray{(2.108)}
    \\
    \multirow{2}{*}{SfC-NeRF$_{-\text{APL}}$}
    & 0.845 & 0.840
    & 0.783 & 0.788
    & 0.805 & 0.786
    & 0.583 & 0.580
    & 0.754 & 0.749
    \\
    & \gray{(1.069)} & \gray{(2.885)}
    & \gray{(1.083)} & \gray{(2.943)}
    & \gray{(0.934)} & \gray{(1.764)}
    & \gray{(0.883)} & \gray{(1.750)}
    & \gray{(0.992)} & \gray{(2.335)}
    \\
    \multirow{2}{*}{SfC-NeRF$_{-\text{APT}}$}
    & 0.624 & 0.631
    & \third{0.428} & 0.604
    & 0.384 & 0.461
    & 0.464 & 0.514
    & 0.475 & 0.553
    \\
    & \gray{(0.588)} & \gray{(1.920)}
    & \gray{(0.586)} & \gray{(1.486)}
    & \gray{(0.579)} & \gray{(1.196)}
    & \gray{(0.646)} & \gray{(1.305)}
    & \gray{(0.600)} & \gray{(1.477)}
    \\
    \multirow{2}{*}{SfC-NeRF$_{-\text{key}}$}
    & \second{0.308} & \first{0.307}
    & \second{0.296} & \second{0.326}
    & \second{0.307} & \second{0.306}
    & \second{0.313} & \second{0.343}
    & \second{0.306} & \second{0.321}
    \\
    & \gray{(0.372)} & \gray{(1.854)}
    & \gray{(0.396)} & \gray{(1.746)}
    & \gray{(0.387)} & \gray{(1.291)}
    & \gray{(0.389)} & \gray{(1.105)}
    & \gray{(0.386)} & \gray{(1.499)}
    \\
    \multirow{2}{*}{SfC-NeRF$_{-\text{VA}}$}
    & 0.542 & 0.611
    & 0.596 & 0.767
    & \third{0.333} & \third{0.389}
    & \third{0.385} & \third{0.421}
    & \third{0.464} & 0.547
    \\
    & \gray{(0.639)} & \gray{(2.304)}
    & \gray{(0.757)} & \gray{(2.265)}
    & \gray{(0.445)} & \gray{(1.338)}
    & \gray{(0.549)} & \gray{(1.339)}
    & \gray{(0.597)} & \gray{(1.811)}
    \\ \midrule
    \multirow{2}{*}{\textstrong{SfC-NeRF}}
    & \first{0.303} & \second{0.308}
    & \first{0.258} & \first{0.313}
    & \first{0.274} & \first{0.273}
    & \first{0.291} & \first{0.307}
    & \first{0.281} & \first{0.300}
    \\
    & \gray{(0.367)} & \gray{(1.821)} 
    & \gray{(0.431)} & \gray{(1.647)} 
    & \gray{(0.448)} & \gray{(1.262)} 
    & \gray{(0.417)} & \gray{(1.204)} 
    & \gray{(0.416)} & \gray{(1.483)} 
    \\ \bottomrule
  \end{tabularx}
  \vspace{-2mm}
  \caption{Comparison of CD and ACD ($\times 10^3$$\downarrow$) when the cavity location $l_c$ is varied.
    This is an extended version of Table~\ref{tab:location}.
    For each condition, the left score indicates CD$_{\text{static}}$, the chamfer distance between $\mathcal{P}(t_0)$ and $\hat{\mathcal{P}}(t_0)$ at the first frame, i.e., $t = t_0$, and the right score indicates CD$_{\text{video}}$, the chamfer distance between $\mathcal{P}(t)$ and $\hat{\mathcal{P}}(t)$ averaged over the entire video sequence, i.e., $t \in \{ t_0, \dots, t_{N-1} \}$.
    The gray score in parentheses indicates the ACD.
    For each condition, the left score indicates ACD$_{\text{static}}$, the anti-chamfer distance at the first frame, and the right score indicates ACD$_{\text{video}}$, the anti-chamfer distance averaged over the entire video sequence.
    It is expected that each original CD is smaller than the corresponding ACD.}
  \vspace{-4mm}
  \label{tab:location_video}
\end{table*}

\subsection{Evaluation from multiple perspectives}
\label{subsec:evaluation_multiple_perspectives}

\subsubsection{Evaluation through video sequences}
\label{subsubsec:evaluation_videos}

In the main experiments, we evaluated the models using the chamfer distance between the ground-truth particles $\hat{\mathcal{P}}^P(t_0)$ and estimated particles $\mathcal{P}^P(t_0)$ in the \textit{first frame} of the video sequence, i.e., at $t = t_0$.
For the multidimensional analysis, we investigated the chamfer distance between the ground-truth particles $\hat{\mathcal{P}}^P(t)$ and estimated particles $\mathcal{P}^P(t)$, averaged over the \textit{entire video sequence}, i.e., $t \in \{ t_0, \dots, t_{N-1} \}$.
For clarity, we refer to the former (chamfer distance for the first static frame) as \textit{CD$_{\text{static}}$} and the latter (chamfer distance for the entire video sequence) as \textit{CD$_{\text{video}}$}.
In the evaluation of the influence of cavity location (Section~\ref{subsec:experiment2}), we introduce anti-chamfer distance, which is the chamfer distance between the predicted particles $\mathcal{P}^P(t_0)$ and ground-truth particles $\tilde{\mathcal{P}}^{P}(t_0)$, where the cavity is placed on the opposite side, in the \textit{first frame} of the video sequence to evaluate how well the cavity location is captured.
For further analysis, we calculated and averaged similar scores for the \textit{entire video sequence}.
For clarity, we refer to the former (anti-chamfer distance for the first static frame) as \textit{ACD$_{\text{static}}$} and the latter (anti-chamfer distance for the entire video sequence) as \textit{ACD$_{\text{video}}$}.

\smallskip\noindent
\textbf{Results.}
Table~\ref{tab:size_video} summarizes the results when the cavity size $s_c$ is varied, and Table~\ref{tab:location_video} summarizes the results when the cavity location $l_c$ is varied.
Our findings are fourfold:

\smallskip\noindent
\textit{(1) CD$_{\text{static}}$ vs. CD$_{\text{video}}$.}
The relative values of CD$_{\text{static}}$ and CD$_{\text{video}}$ vary across different cases.
When calculating CD$_{\text{static}}$ in the first frame, the locations of the ground-truth and synthesized objects were well aligned, allowing us to focus on the differences in shapes.
In contrast, when calculating CD$_{\text{video}}$ for the entire video sequence, we must consider not only the differences in shapes but also the differences in absolute locations.
Misalignments accumulate over time because the locations must vary within the allowance of the physical constraints via DiffMPM~\cite{YHuICLR2020}.
Because the objective of this study was to correctly predict the shape rather than the location, CD$_{\text{static}}$ is a more valid evaluation than CD$_{\text{video}}$ for this purpose.

\smallskip\noindent
\textit{(2) Comparison of CD$_{\text{static}}$ and CD$_{\text{video}}$ among models.}
Although there was some variation in the superiority of the models depending on the metric used, the general trend remained consistent: SfC-NeRF achieved the best score in most cases.
The two exceptions are CD$_{\text{video}}$ for $s_c = 0$ in Table~\ref{tab:size_video} and CD$_{\text{video}}$ for $l_c = \text{left}$ in Table~\ref{tab:location_video}. However, the difference from the best score is small (less than $0.002$).
These results validate the effectiveness of the proposed method compared to the baseline and ablated models according to both metrics.

\smallskip\noindent
\textit{(3) ACD$_{\text{static}}$ vs. ACD$_{\text{video}}$.}
Comparing ACD$_{\text{static}}$ with ACD$_{\text{video}}$, ACD$_{\text{static}}$ is smaller than ACD$_{\text{video}}$.
This is because the difference in location gradually increased after the collision when the cavity was located on the opposite side.
As the objective of this study was to correctly predict the shape rather than the location, ACD$_{\text{static}}$ is a more valid evaluation than ACD$_{\text{video}}$ for this purpose.

\smallskip\noindent
\textit{(4) Comparison of CD$_{\text{static}}$ and ACD$_{\text{static}}$ among models.}
When comparing the models, the baselines (i.e., the GO- and LPO-based models) tended to obtain similar CD$_{\text{static}}$ and ACD$_{\text{static}}$ values because they struggled to determine the optimization direction, as shown in Figures~\ref{fig:results_sphere}--\ref{fig:results_diamond}.
In contrast, the proposed models (i.e., the SfC-NeRF-based models, including the ablated models) tended to obtain a smaller CD$_{\text{static}}$ than ACD$_{\text{static}}$.
These results indicate that the proposed models effectively capture the positional bias of the cavity.
Notably, a larger ACD$_{\text{static}}$ does not indicate better performance unless CD$_{\text{static}}$ is adequately small because it is possible to increase ACD$_{\text{static}}$ while sacrificing CD$_{\text{static}}$.

\subsubsection{Evaluation per external shape}
\label{subsubsec:evaluation_per_shape}

In Experiments I (Section~\ref{subsec:experiment1}) and II (Section~\ref{subsec:experiment2}), we reported the scores averaged over external shapes (i.e., sphere, cube, bicone, cylinder, and diamond objects).
For a different evaluation perspective, this appendix presents the scores for each external shape, averaged over other conditions, i.e., either $s_c \in \{ 0, (\frac{1}{2})^3, (\frac{2}{3})^3, (\frac{3}{4})^3 \}$ or $l_c \in \{ \text{left}, \text{right}, \text{up}, \text{down} \}$.

\smallskip\noindent
\textbf{Results.}
Table~\ref{tab:size_per_shape} summarizes the results when the cavity size $s_c$ is varied (related to the results in Table~\ref{tab:size}), and Table~\ref{tab:location_per_shape} summarizes the results when the cavity location $l_c$ is varied (related to the results in Table~\ref{tab:location}).
Although the scores were affected by the external shape, the same trends observed previously regarding the superiority or inferiority of the models were maintained.
In particular, SfC-NeRF outperformed both the baseline and ablated models in most cases.

\begin{table}[t]
  \centering
  \footnotesize
  \setlength{\tabcolsep}{0.1pt}
  \begin{tabularx}{\columnwidth}{lCCCCC}
    \toprule
    & Sphere & Cube & Bicone & Cylinder & Diamond
    \\ \midrule
    Static
    & 0.897
    & 0.612
    & 0.724
    & 0.697
    & 0.671
    \\ \midrule
    GO
    & 0.889
    & 0.637
    & 0.704
    & 0.756
    & 0.663
    \\
    GO$_{\text{mass}}$
    & 0.934
    & 1.345
    & 0.760
    & 1.218
    & 0.663
    \\
    LPO
    & 0.774
    & 0.564
    & 0.639
    & 0.678
    & 0.622
    \\
    LPO$_{\text{mass}}$
    & 0.796
    & 0.605
    & 0.656
    & 0.726
    & 0.622
    \\ \midrule
    SfC-NeRF$_{-\text{mass}}$
    & 0.561
    & 0.500
    & 0.455
    & \third{0.447}
    & 0.553
    \\
    SfC-NeRF$_{-\text{APL}}$
    & 0.303
    & 1.082
    & 0.579
    & 0.885
    & 0.591
    \\
    SfC-NeRF$_{-\text{APT}}$
    & 0.178
    & 0.375
    & \third{0.286}
    & 0.502
    & 0.331
    \\
    SfC-NeRF$_{-\text{key}}$
    & \second{0.081}
    & \second{0.173}
    & \second{0.159}
    & \second{0.288}
    & \second{0.230}
    \\
    SfC-NeRF$_{-\text{VA}}$
    & \third{0.113}
    & \third{0.279}
    & 0.363
    & 0.558
    & \third{0.268}
    \\ \midrule
    \textstrong{SfC-NeRF}
    & \first{0.067}
    & \first{0.163}
    & \first{0.138}
    & \first{0.264}
    & \first{0.193}
    \\ \bottomrule
  \end{tabularx}
  \vspace{-2mm}
  \caption{Comparison of CD ($\times 10^3$$\downarrow$) when the cavity size $s_c$ is varied.
    The scores were averaged over cavity sizes.}
  \label{tab:size_per_shape}
\end{table}

\begin{table}[t]
  \centering
  \footnotesize
  \setlength{\tabcolsep}{0.1pt}
  \begin{tabularx}{\columnwidth}{lCCCCC}
    \toprule
    & Sphere & Cube & Bicone & Cylinder & Diamond
    \\ \midrule
    Static
    & 1.006
    & 0.719
    & 0.824
    & 0.818
    & 0.772
    \\ \midrule
    GO
    & 0.991
    & 0.809
    & 0.847
    & 0.898
    & 0.799
    \\
    GO$_{\text{mass}}$
    & 1.065
    & 1.528
    & 0.934
    & 1.332
    & 1.125
    \\
    LPO
    & 0.954
    & 0.673
    & 0.764
    & 0.804
    & 0.701
    \\
    LPO$_{\text{mass}}$
    & 0.980
    & 0.723
    & 0.796
    & 0.845
    & 0.711
    \\ \midrule
    SfC-NeRF$_{-\text{mass}}$
    & 0.695
    & 0.480
    & 0.424
    & 0.595
    & 0.533
    \\
    SfC-NeRF$_{-\text{APL}}$
    & 0.548
    & 1.064
    & 0.373
    & 1.194
    & 0.592
    \\
    SfC-NeRF$_{-\text{APT}}$
    & 0.318
    & 0.502
    & \third{0.374}
    & 0.730
    & 0.451
    \\
    SfC-NeRF$_{-\text{key}}$
    & \second{0.189}
    & \second{0.371}
    & \second{0.235}
    & \second{0.448}
    & \first{0.286}
    \\
    SfC-NeRF$_{-\text{VA}}$
    & \third{0.240}
    & \third{0.418}
    & 0.790
    & \third{0.534}
    & \third{0.338}
    \\ \midrule
    \multirow{2}{*}{\textstrong{SfC-NeRF}}
    & \first{0.152}
    & \first{0.342}
    & \first{0.231}
    & \first{0.393}
    & \second{0.289}
    \\
    & \gray{(0.417)}
    & \gray{(0.386)}
    & \gray{(0.365)}
    & \gray{(0.491)}
    & \gray{(0.420)}
    \\ \bottomrule
  \end{tabularx}
  \vspace{-2mm}
  \caption{Comparison of CD ($\times 10^3$$\downarrow$) when the cavity location $l_c$ is varied.
    The scores were averaged over cavity locations.
    The gray score in parentheses indicates ACD ($\times 10^3$).
    It is expected that the original CD is smaller than this.}
  \vspace{-4mm}
  \label{tab:location_per_shape}
\end{table}

\subsection{Possible challenges with real data}
\label{subsec:challenge_real}

As discussed in Section~\ref{sec:discussion}, because \textit{SfC} is a novel task, this study focused on evaluating its fundamental performance using simulation data, leaving validation with real data a challenge for future research.
However, it is both feasible and important to discuss the potential challenges associated with real data, which we address in this appendix.
Three potential challenges are outlined below:

\smallskip\noindent
\textit{(1) Difficulty in accurately estimating external structures.}
Although significant progress has been made in the estimation of 3D external structures in recent years, it is not yet possible to accurately estimate them for all objects in all situations.
The proposed method assumes that the external structure learned in the first frame of the video sequence is accurate.
Therefore, if this estimation fails, the overall performance is degraded.
We believe that incorporating the concept of a physics-informed model, particularly in challenging scenarios (e.g., sparse views), such as Lagrangian particle optimization~\cite{TKanekoCVPR2024}, could provide a solution to this issue.

\smallskip\noindent
\textit{(2) Gap between real physics and the physics used in the simulation.}
Despite recent advancements in physical simulation models, discrepancies between real-world physics and the physics underlying the simulation still persist.
We believe that refining the proposed method alongside physics-informed models (e.g., those discussed in Section~\ref{sec:related_work}) could help alleviate this problem.

\smallskip\noindent
\textit{(3) Difficulty in accurately estimating physical properties.}
As mentioned in Section~\ref{subsec:problem}, we address \textit{SfC} under the assumption that the ground-truth physical properties are available in advance to mitigate the chicken-and-egg problem between the physical properties and internal structures.
This assumption is reasonable if the material can be identified; however, obtaining perfectly accurate values for physical properties in real-world scenarios is challenging.
Although the issue of solving the chicken-and-egg problem remains, an appearance-based physical property estimation method has already been proposed (e.g., PAC-NeRF~\cite{XLiICLR2023}).
Combining the proposed approach with previous methods for the simultaneous optimization of physical properties and internal structures is an exciting direction for future research.

\section{Qualitative results}
\label{sec:qualitative_results}

This appendix presents the qualitative results.
The corresponding demonstration videos are available at \url{https://www.kecl.ntt.co.jp/people/kaneko.takuhiro/projects/sfc/}.

\subsection{Qualitative results of Experiments I and II}
\label{sec:qualitative_results_experiment1_2}

\begin{figure*}[h]
  \centering
  \includegraphics[width=\linewidth]{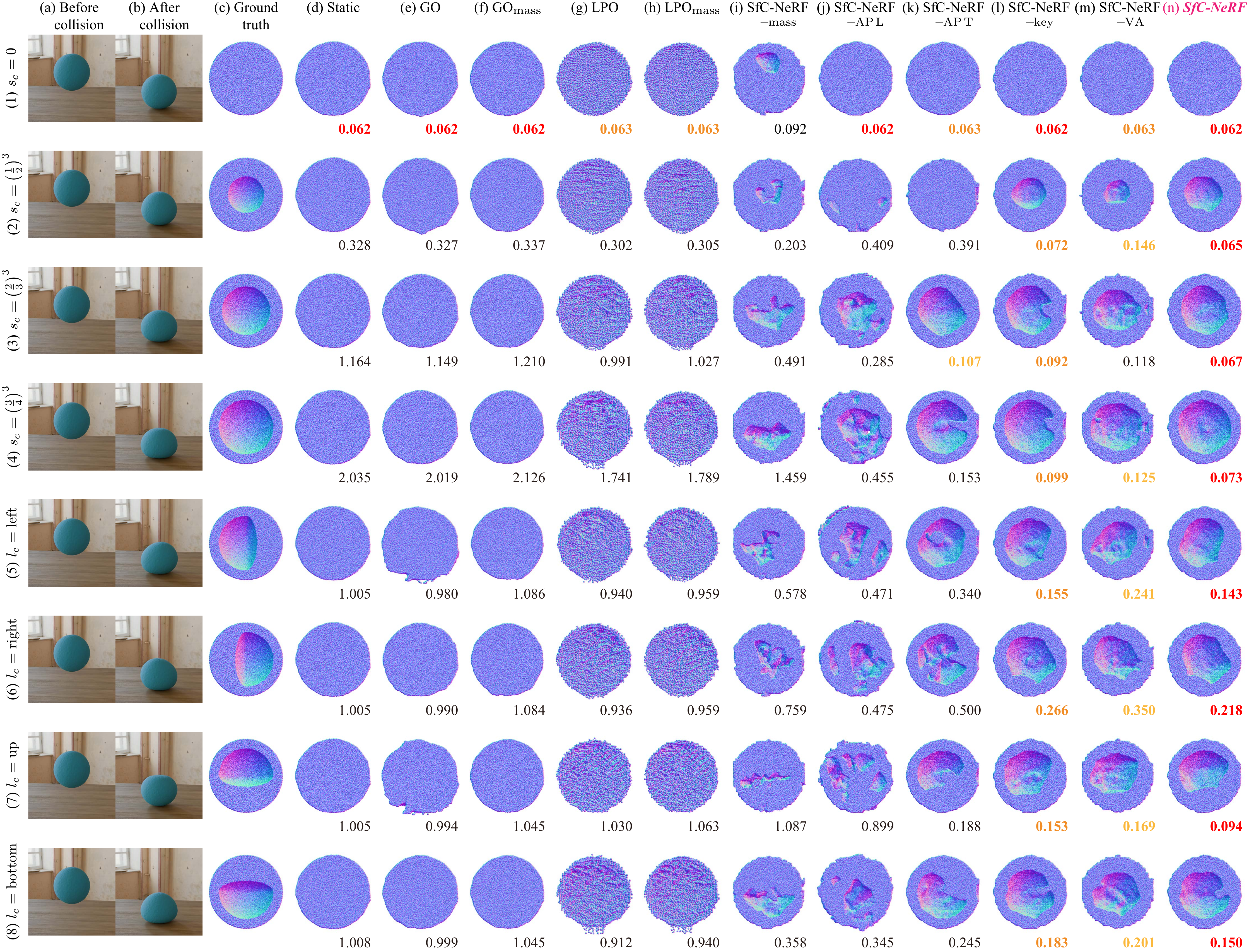}
  \caption{Comparison of learned internal structures for \textit{sphere} objects.
    \textit{(a) and (b) Examples of training images.}
    The images are zoomed in for easy viewing.
    \textit{(a) Examples of training images \textbf{before} collision.}
    As shown in this column, the appearances of the objects are the same across all scenes (1)--(8).
    Consequently, it is difficult to distinguish the internal structures based solely on these appearances.
    \textit{(b) Examples of training images \textbf{after} collision.}
    To overcome the difficulty mentioned above, we address \textit{SfC}, in which we aim to identify the internal structures based on appearance changes before and after collision, as shown in (a) and (b).
    \textit{(c)--(n) Internal structures visualized through cross-sectional views perpendicular to the ground.}
    In (d)--(n), the score below each image indicates CD ($\times 10^3$$\downarrow$).
    \textit{(c) Ground-truth internal structures.}
    As shown in this column, although the external appearances are the same in (a), the internal structures are different.
    \textit{(d) Internal structures learned from the first frames of the video sequences.}
    The same internal structures (i.e., filled objects) were learned because the appearances were the same before the collision (a).
    \textit{(e)--(h) Internal structures learned using the baselines (GO- and LPO-based models).}
    These models struggled to determine optimal learning directions.
    \textit{(i)--(m) Internal structures learned using the ablated models.}
    The ablated models are insufficient to prevent convergence to improper solutions.
    \textit{(n) Internal structures learned using SfC-NeRF (full model).}
    The full model overcomes the above drawbacks and achieved the best CD.}
  \label{fig:results_sphere}
\end{figure*}

\begin{figure*}[t]
  \centering
  \includegraphics[width=\linewidth]{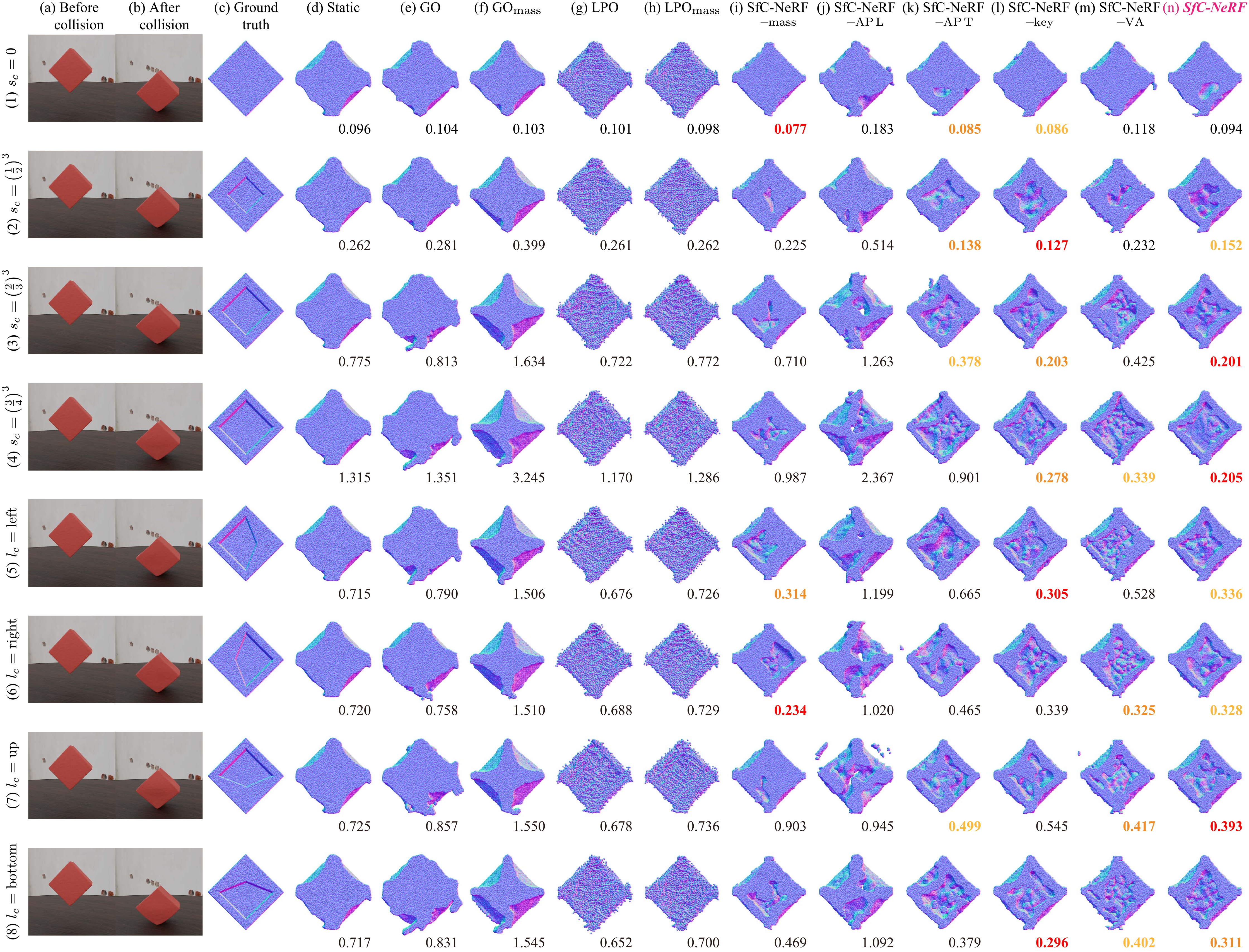}
  \caption{Comparison of learned internal structures for \textit{cube} objects.
    The view in the figure is the same as that of Figure~\ref{fig:results_sphere}.}
  \label{fig:results_cube}
\end{figure*}

\begin{figure*}[t]
  \centering
  \includegraphics[width=\linewidth]{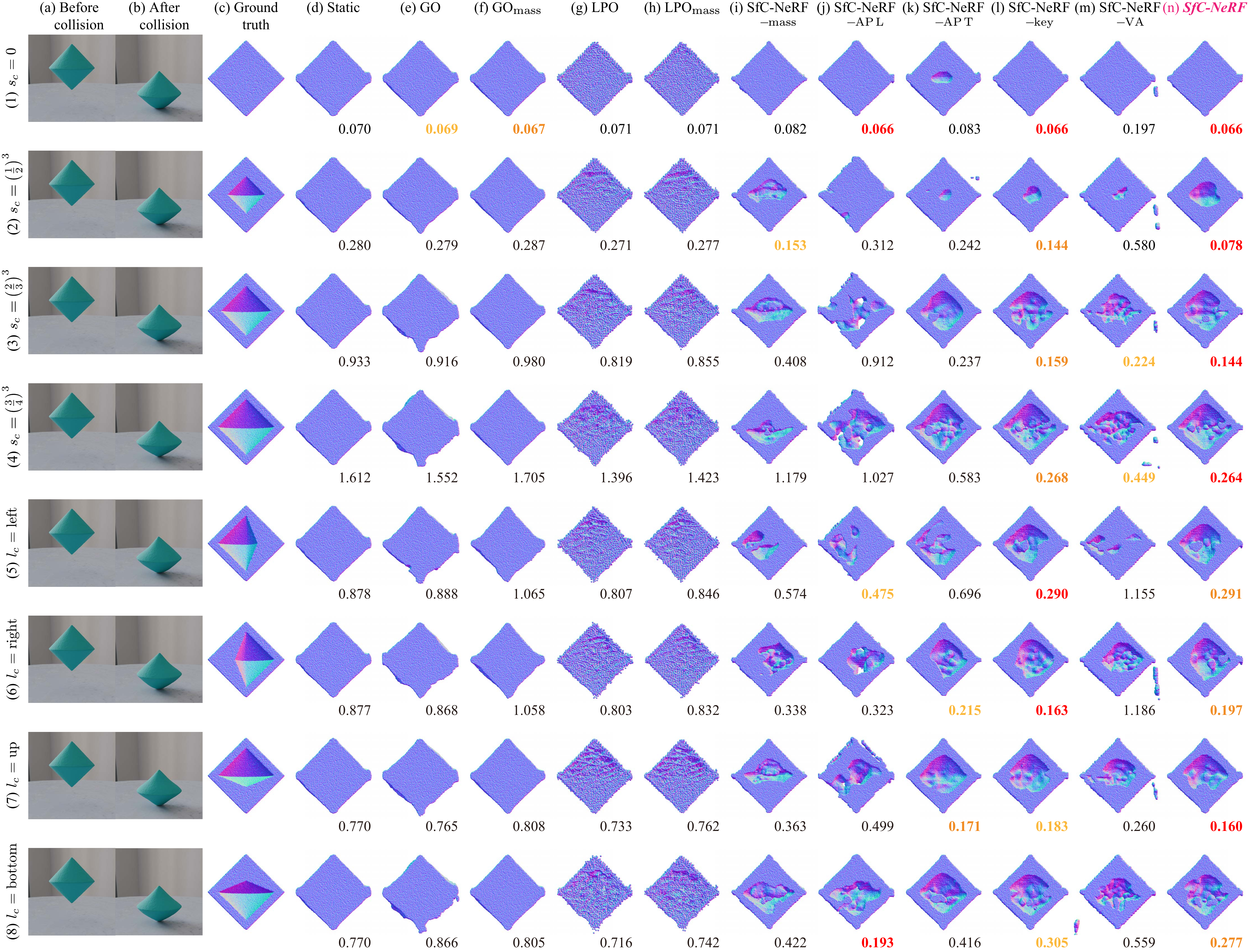}
  \caption{Comparison of learned internal structures for \textit{bicone} objects.
    The view in the figure is the same as that of Figure~\ref{fig:results_sphere}.}
  \label{fig:results_bicone}
\end{figure*}

\begin{figure*}[t]
  \centering
  \includegraphics[width=\linewidth]{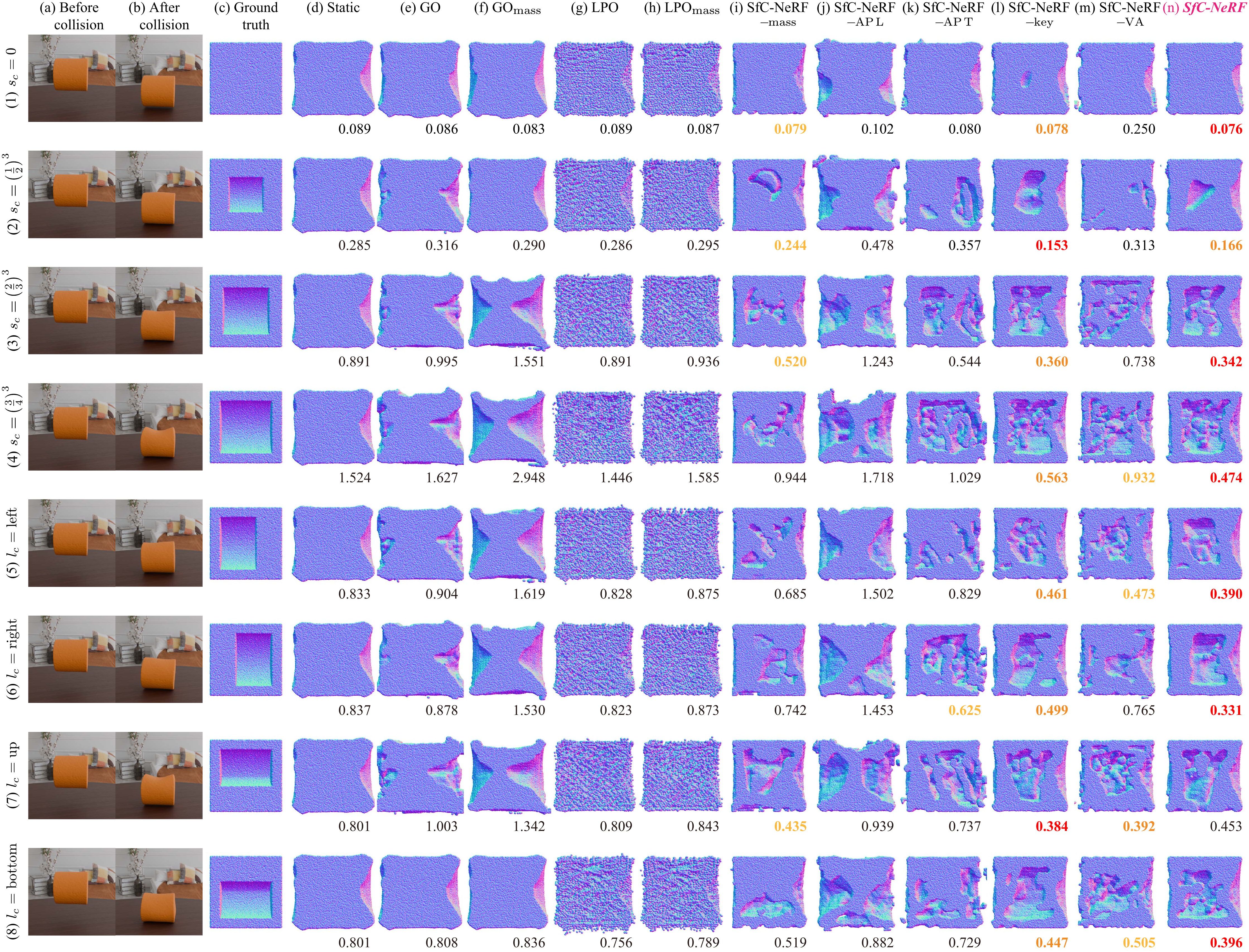}
  \caption{Comparison of learned internal structures for \textit{cylinder} objects.
    The view in the figure is the same as that of Figure~\ref{fig:results_sphere}.}
  \label{fig:results_cylinder}
\end{figure*}

\begin{figure*}[t]
  \centering
  \includegraphics[width=\linewidth]{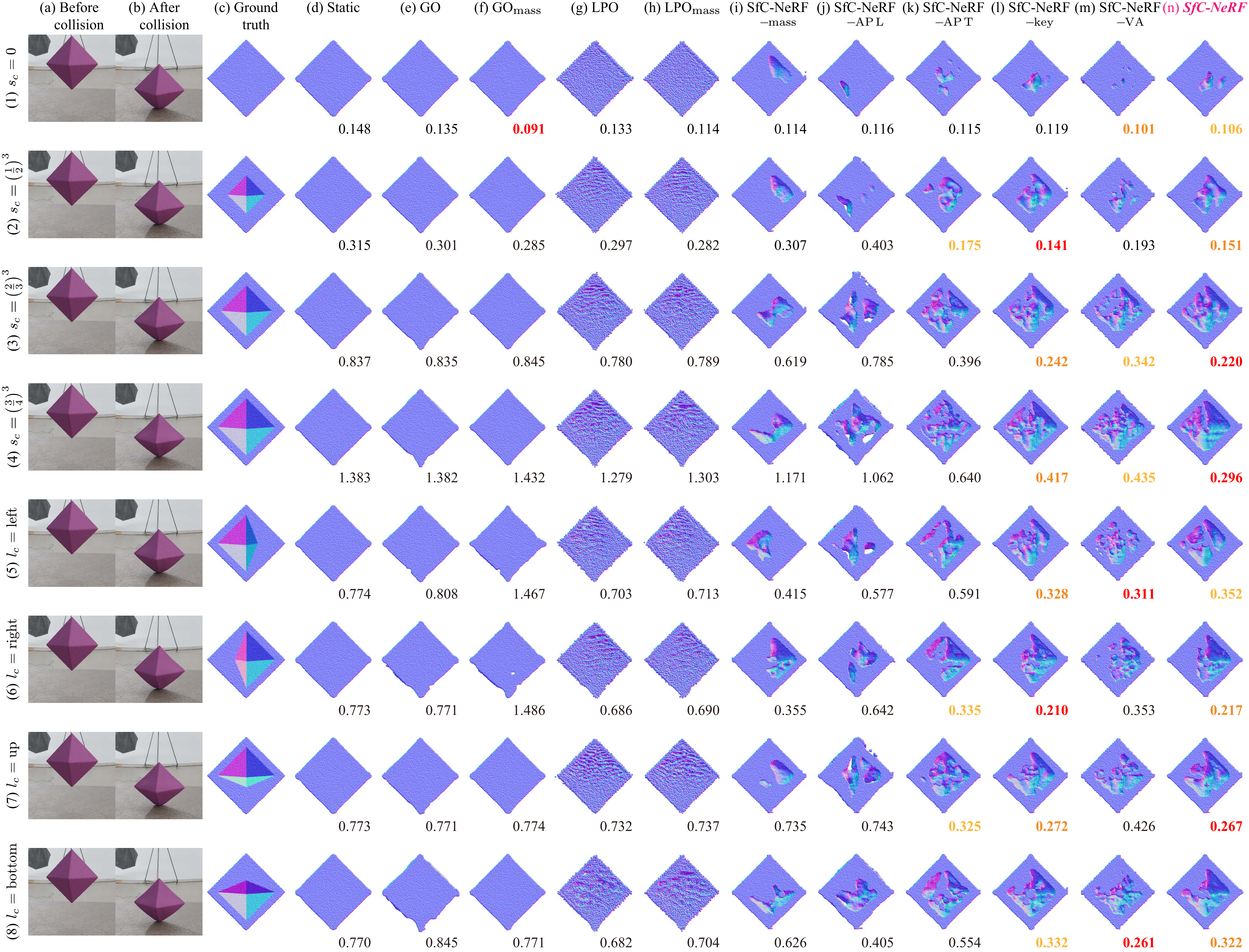}
  \caption{Comparison of learned internal structures for \textit{diamond} objects.
    The view in the figure is the same as that of Figure~\ref{fig:results_sphere}.}
  \label{fig:results_diamond}
\end{figure*}

We provide the qualitative results of Experiments I (Section~\ref{subsec:experiment1}) and II (Section~\ref{subsec:experiment2}) in Figures~\ref{fig:results_sphere}--\ref{fig:results_diamond}.

\subsection{Qualitative results of Experiment III}
\label{subsec:qualitative_results_experiment3}

\begin{figure*}[t]
  \centering
  \includegraphics[width=\linewidth]{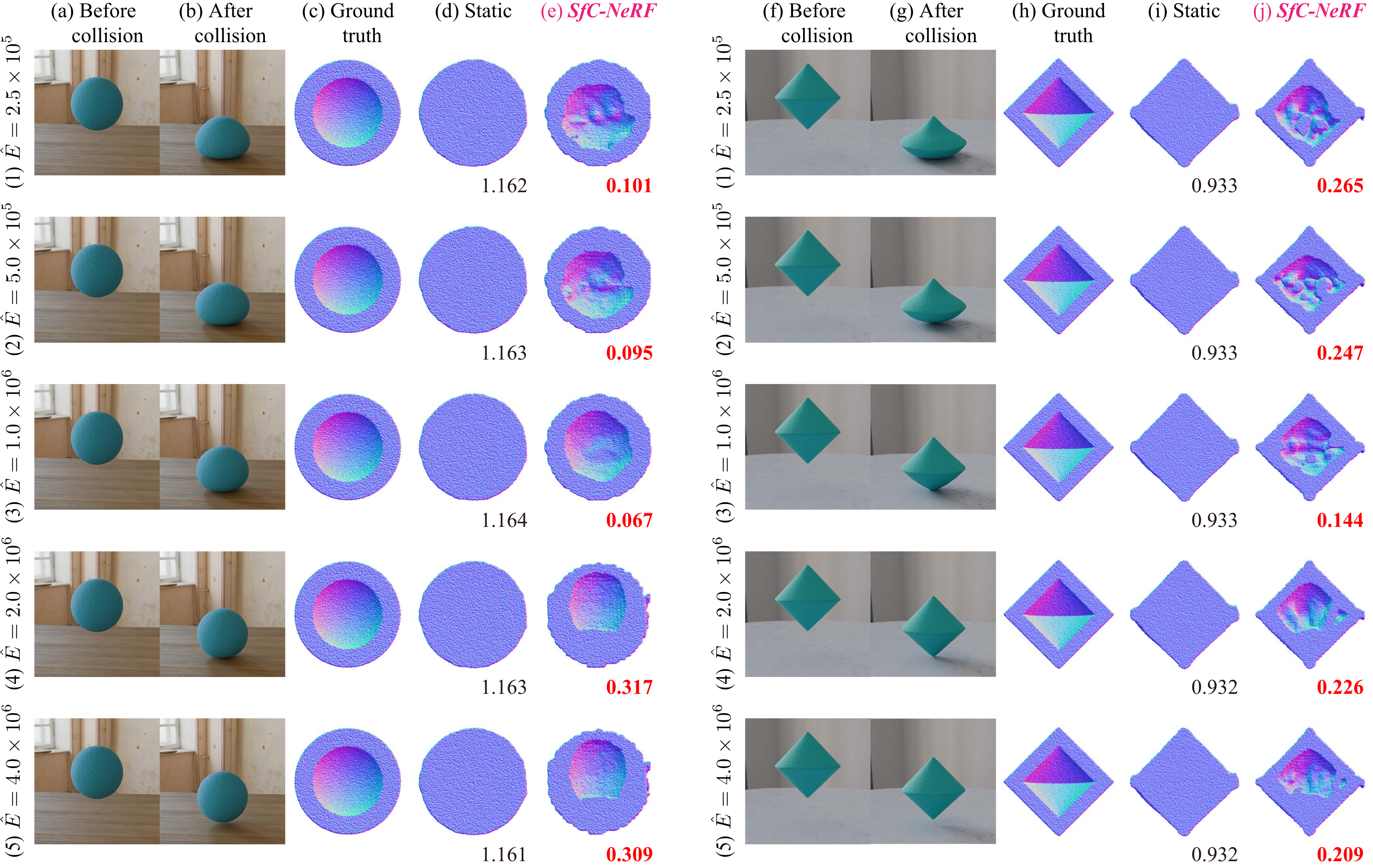}
  \caption{Comparison of learned internal structures for \textit{sphere} objects (left) and \textit{bicone} objects (right) when Young's modulus $\hat{E}$ is varied.
    Young's modulus is a measure of elasticity and quantifies tensile or compressive stiffness when force is applied.
    Here, we discuss the results for the sphere objects because the same tendencies were observed for the bicone objects.
    As shown in (a) and (c), the external appearances before collision (a) and internal structures (c) are the same in all cases (1)--(5).
    However, as shown in (b), the shapes after collision differ because of variations in Young's modulus $\hat{E} \in \{ 2.5 \times 10^5, 5.0 \times 10^5, 1.0 \times 10^6, 2.0 \times 10^6, 4.0 \times 10^6 \}$.
    In particular, as Young's modulus increases from top to bottom, the object becomes stiffer, and the amount of shape change decreases.
    In the Static model (b), the internal structure was learned from the first frame, which looks the same in all cases.
    As a result, the same internal structure was learned across all variations.
    In contrast, in SfC-NeRF (e), the internal structure was learned using video sequences with different appearances.
    In this example, the same internal structure is expected to be learned in all cases.
    However, the varying appearances after collision (b), which provide a clue for solving the problem, lead to different outcomes.
    As shown in (1)(b) and (2)(b), when the object is soft, it deforms significantly after collision.
    This makes it difficult to capture the internal structure consistently, as shown in (1)(e) and (2)(e).
    In contrast, as shown in (4)(b) and (5)(b), when the object is stiffer, the shape change is limited.
    This narrows the range within which internal structures can be estimated, as shown in (4)(e) and (5)(e).
    Because \textit{SfC} is an ill-posed problem with multiple possible solutions, the obtained results are considered reasonable.
    However, further improvement remains a topic for future work.}
  \label{fig:results_young}
\end{figure*}

\begin{figure*}[t]
  \centering
  \includegraphics[width=\linewidth]{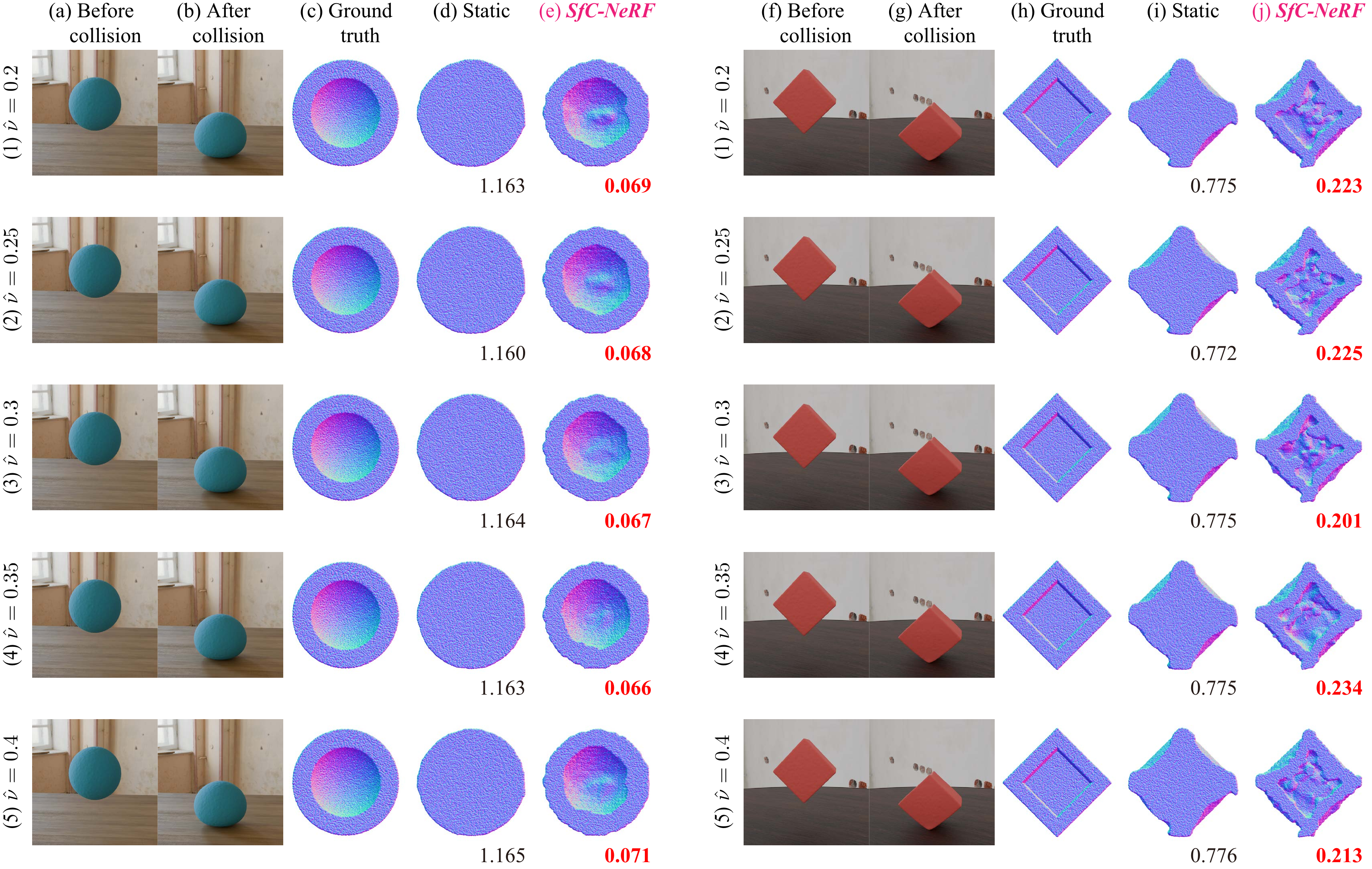}
  \caption{Comparison of learned internal structures for \textit{sphere} objects (left) and \textit{cube} objects (right) when Poisson's ratio $\hat{\nu}$ is varied.
    Poisson's ratio is a measure of the Poisson effect and quantifies how much a material deforms in a direction perpendicular to the direction in which force is applied.
    We varied Poisson's ratio $\hat{\nu}$ within the range of values commonly observed in real materials, i.e., $\hat{\nu} \in \{ 0.2, 0.25, 0.3, 0.35, 0.4 \}$.
    As shown in (b) and (g), this physical property does not significantly affect the appearance after the collision compared to the results when Young's modulus is varied (Figure~\ref{fig:results_young}).
    As a result, the learned internal structures are almost identical, as shown in (e) and (j).}
  \label{fig:results_poisson}
\end{figure*}

\begin{figure*}[t]
  \centering
  \includegraphics[width=\linewidth]{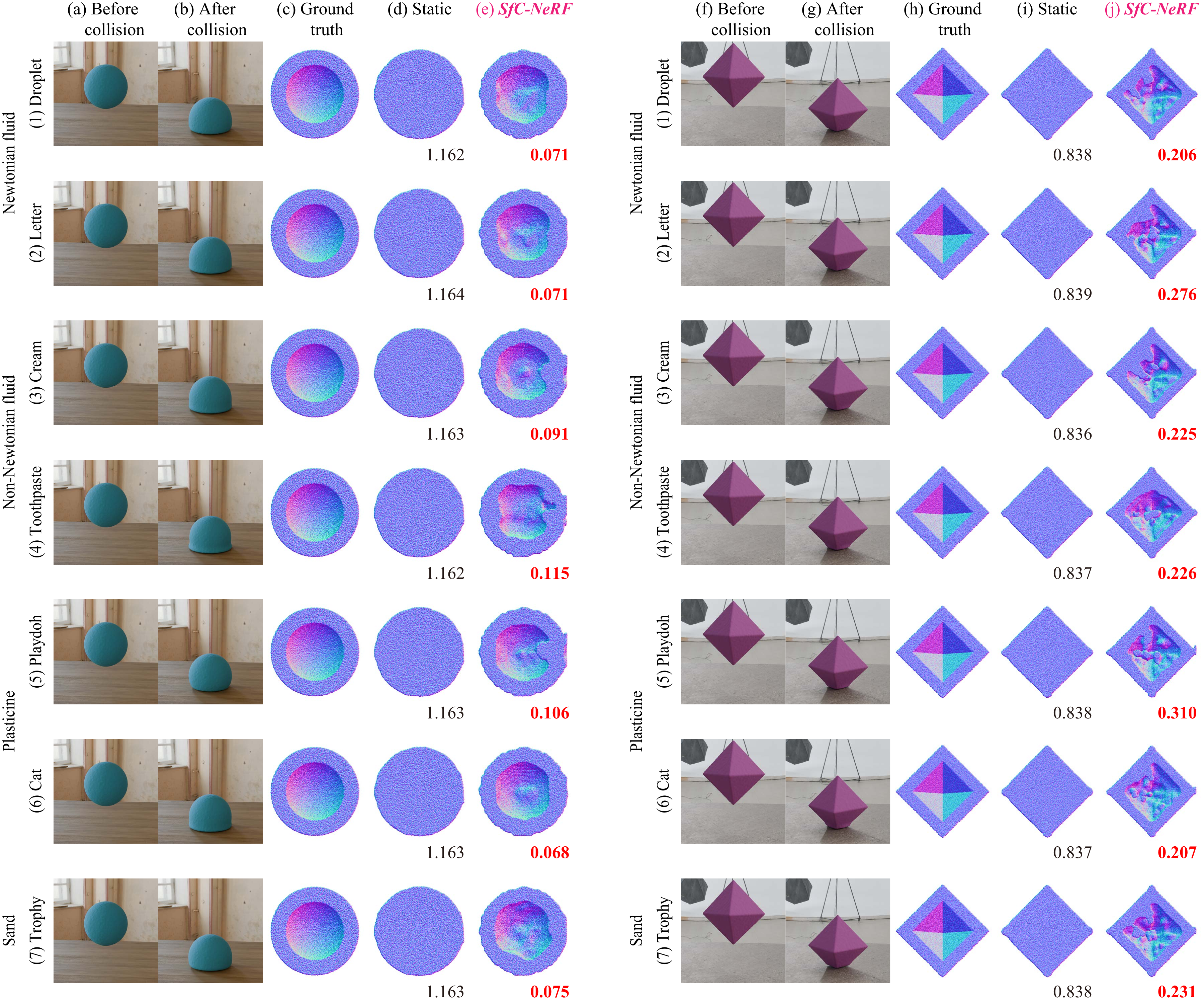}
  \caption{Comparison of learned internal structures for \textit{sphere} objects (left) and \textit{diamond} objects (right) with varying materials.
    The physical properties were based on the PAC-NeRF dataset~\cite{XLiICLR2023}.
    Specifically:
    (1) Newtonian fluid with the ``Droplet'' setting (fluid viscosity $\hat{\mu} = 200$ and bulk modulus $\hat{\kappa} = 10^5$).
    (2) Newtonian fluid with the ``Letter'' setting ($\hat{\mu} = 100$ and $\hat{\kappa} = 10^5$).
    (3) Non-Newtonian fluid with the ``Cream'' setting (shear modulus $\hat{\mu} = 10^4$, bulk modulus $\hat{\kappa} = 10^6$, yield stress $\hat{\tau}_Y = 3 \times 10^3$, and plasticity viscosity $\hat{\eta} = 10$).
    (4) Non-Newtonian fluid with the ``Toothpaste'' setting ($\hat{\mu} = 5 \times 10^3$, $\hat{\kappa} = 10^5$, $\hat{\tau}_Y = 200$, and $\hat{\eta} = 10$).
    (5) Plasticine with the ``Playdoh'' setting (Young's modulus $\hat{E} = 2 \times 10^6$, Poisson's ratio $\hat{\nu} = 0.3$, and yield stress $\hat{\tau}_Y = 1.54 \times 10^4$).
    (6) Plasticine with the ``Cat'' setting ($\hat{E} = 10^6$, $\hat{\nu} = 0.3$, and $\hat{\tau}_Y = 3.85 \times 10^3$).
    (7) Sand with the ``Trophy'' setting ($\hat{\theta}_{fric} = 40\tcdegree$).
    These results demonstrate that \textit{SfC-NeRF} ((e) and (j)) improves structure estimation compared to Static ((d) and (i)), regardless of the material.
    However, the improvement rate depends on the material.
    As an initial approach to address \textit{SfC}, we proposed a general-purpose method.
    However, it would be interesting to develop methods specifically tailored to individual materials in future work.}
  \label{fig:results_material}
\end{figure*}

We provide the qualitative results of Experiment III (Section~\ref{subsec:experiment3}) in Figures~\ref{fig:results_young}--\ref{fig:results_material}.

\subsection{Qualitative results of Experiment IV}
\label{subsec:qualitative_results_experiment4}

\begin{figure*}[t]
  \centering
  \includegraphics[width=\linewidth]{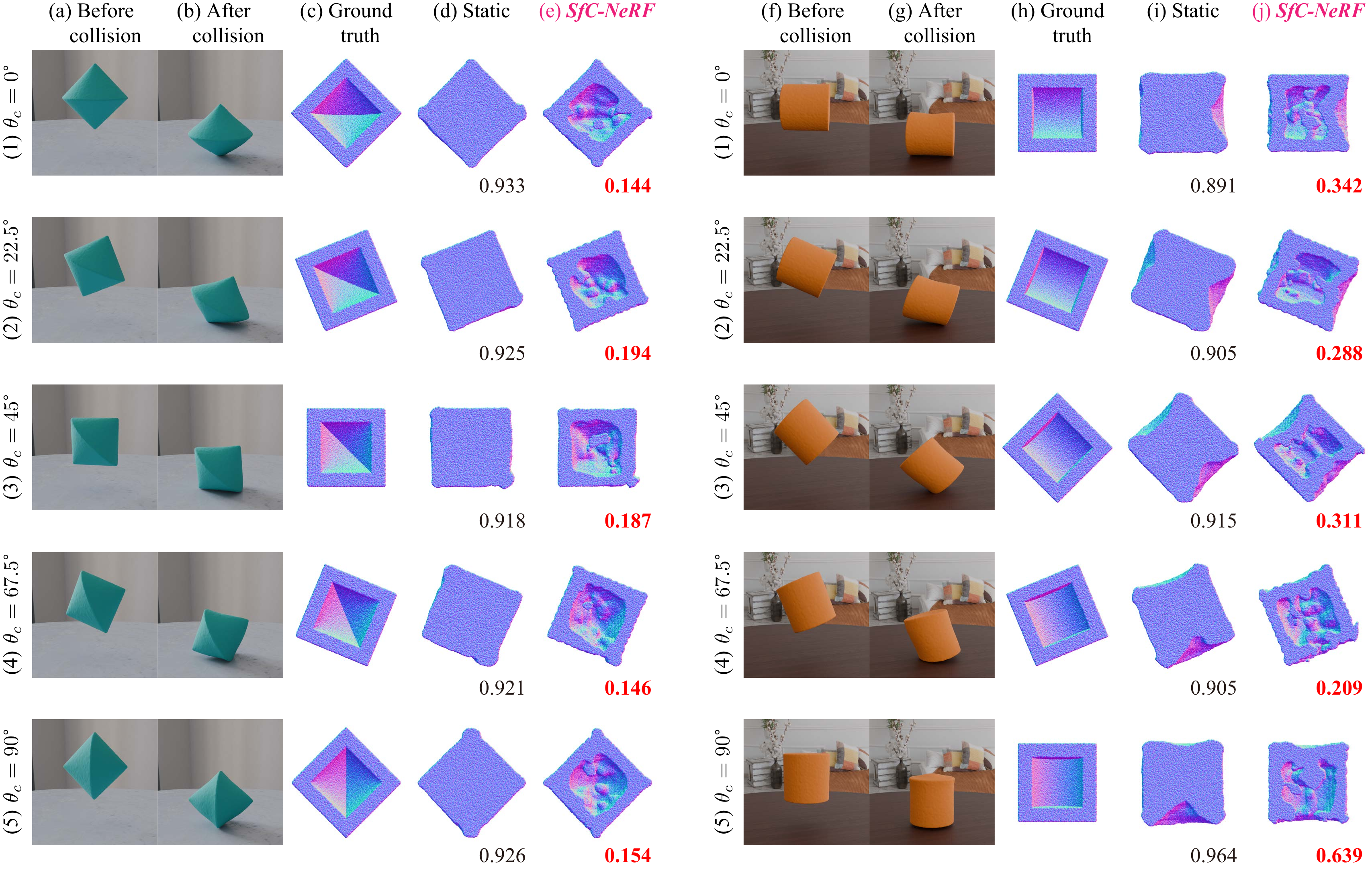}
  \caption{Comparison of learned internal structures for \textit{bicone} objects (left) and \textit{cylinder} objects (right) when collision angle $\theta_c$ is varied.
    We varied collision angle $\theta_c \in \{ 0\tcdegree, 22.5\tcdegree, 45\tcdegree, 67.5\tcdegree, 90\tcdegree \}$.
    We found that the effect of collision angle on the estimation of the internal structure depends on the object shape.
    (a)--(e) In the case of an object such as \textit{bicone}, where the object is entirely visible regardless of the collision angle, the estimation performance remains relatively stable across different collision angles.
    (f)--(j) In contrast, in the case of an object, such as \textit{cylinder}, where the visible area varies greatly depending on the collision angle, the estimation performance also changes with the collision angle.
    For example, in (5)(g), the bottom of the object is not visible when it collides with the ground.
    As a result, a hole is generated at the bottom of the object in (5)(j).
    This issue may be alleviated by improving camera placement.
    Other possible factors that affect estimation performance are discussed in Appendix~\ref{subsubsec:experiment4}.}
  \label{fig:results_angle}
\end{figure*}

We provide the qualitative results of Experiment IV (Appendix~\ref{subsubsec:experiment4}) in Figure~\ref{fig:results_angle}.

\clearpage
\clearpage
\section{Implementation details}
\label{sec:implementation}

\subsection{Dataset}
\label{subsec:dataset}

Because \textit{SfC} is a new task and no established dataset is available, we created a new dataset called the \textit{SfC dataset} based on the protocol of PAC-NeRF~\cite{XLiICLR2023}, which is a pioneering study on geometry-agnostic system identification.
In the main experiments presented in Section~\ref{sec:experiments}, we prepared $115$ objects by changing their external shapes, internal structures, and materials.
Figure~\ref{fig:dataset} shows examples of the data in this dataset.
First, we prepared five external shapes: \textit{sphere}, \textit{cube}, \textit{bicone}, \textit{cylinder}, and \textit{diamond}.
Regarding the internal structure and material, we set the default values as follows:
the cavity size rate for the filled object, $s_c$, was set to $(\frac{2}{3})^3$,
the cavity location, $l_c$, was set to the center, and
the material was defined as an elastic material with Young's modulus $\hat{E} = 10^6$ and Poisson's ratio $\hat{\nu} = 0.3$.
Under these default properties, one of them was changed as follows:

\smallskip\noindent
\textit{(a) Three different sized cavities}: $s_c \in \{ 0, (\frac{1}{2})^3, (\frac{3}{4})^3 \}$.

\smallskip\noindent
\textit{(b) Four different locations of cavities}: the center $l_c$ is moved $\{ \text{up}, \text{down}, \text{left}, \text{right} \}$.

\smallskip\noindent
\textit{(c-1) Eight different elastic materials}: those with four different Young's moduli $\hat{E} \in \{ 2.5 \times 10^5, 5 \times 10^5, 2 \times 10^6, 4 \times 10^6\}$ and four different Poisson's ratios $\hat{\nu} \in \{ 0.2, 0.25, 0.35, 0.4 \}$.

\smallskip\noindent
\textit{(c-2) Seven different materials}: two Newtonian fluids, two non-Newtonian fluids, two plasticines, and one sand.
Their physical properties were derived from the PAC-NeRF dataset~\cite{XLiICLR2023}.
Specifically, the two Newtonian fluids included one with the ``Droplet'' setting (fluid viscosity $\hat{\mu} = 200$ and bulk modulus $\hat{\kappa} = 10^5$) and one with the ``Letter'' setting ($\hat{\mu} = 100$ and $\hat{\kappa} = 10^5$).
The two non-Newtonian fluids included one with the ``Cream'' setting (shear modulus $\hat{\mu} = 10^4$, bulk modulus $\hat{\kappa} = 10^6$, yield stress $\hat{\tau}_Y = 3 \times 10^3$, and plasticity viscosity $\hat{\eta} = 10$) and one with the ``Toothpaste'' setting ($\hat{\mu} = 5 \times 10^3$, $\hat{\kappa} = 10^5$, $\hat{\tau}_Y = 200$, and $\hat{\eta} = 10$).
The two plasticines included one with the ``Playdoh'' setting (Young's modulus $\hat{E} = 2 \times 10^6$, Poisson's ratio $\hat{\nu} = 0.3$, and yield stress $\hat{\tau}_Y = 1.54 \times 10^4$) and one with the ``Cat'' setting ($\hat{E} = 10^6$, $\hat{\nu} = 0.3$, and $\hat{\tau}_Y = 3.85 \times 10^3$).
The sand had the ``Trophy'' setting ($\hat{\theta}_{fric} = 40\tcdegree$).

\smallskip
Thus, we created $5$ external shapes $\times$ ($1$ default $+$ $3$ sizes $+$ $4$ locations $+$ $(8 + 7)$ materials) $= 115$ objects.

We also prepared $20$ objects for the extended experiments described in Appendix~\ref{subsec:extended_experiments}.
Specifically, we considered four collision angles: $\theta_c \in \{ 22.5\tcdegree, 45\tcdegree, 67.5\tcdegree, 90\tcdegree \}$.
Thus, in this appendix, we created $5$ external shapes $\times$ $4$ collision angles $= 20$ objects.
The total number of objects created in the main text and this appendix is $115 + 20 = 135$.

Following the PAC-NeRF study~\cite{XLiICLR2023}, the ground-truth data were generated using the MLS-MPM simulator~\cite{YHuTOG2018}, where each object fell freely under the influence of gravity and collided with the ground plane.
Images were rendered under various environmental lighting conditions and ground textures using a photorealistic renderer.
Each scene was captured from 11 viewpoints using cameras spaced in the upper hemisphere including an object.

\subsection{Model}
\label{subsec:model}

We implemented the models based on the official PAC-NeRF code~\cite{XLiICLR2023}.\footnote{\url{https://github.com/xuan-li/PAC-NeRF}}
PAC-NeRF represents an Eulerian grid-based scene representation using voxel-based NeRF (specifically, direct voxel grid optimization (DVGO)~\cite{CSunCVPR2022}) and conducts a Lagrangian particle-based differentiable physical simulation using a differentiable MPM simulator (specifically, DiffTaichi~\cite{YHuICLR2020}).
More specifically, DVGO represents a volume density field $\sigma^{G'}$ using a 3D dense voxel grid and represents a color field $\mathbf{c}^{G'}$ using a combination of a 4D dense voxel grid and a two-layer multi-layer perceptron (MLP) with a hidden dimension of $128$.
When the MLP is employed, positional embedding in the viewing direction $\mathbf{d}$ is used as an additional input.
We set the resolutions of $\sigma^{G'}$ and $\mathbf{c}^{G'}$ to match those in PAC-NeRF~\cite{XLiICLR2023}.

\subsection{Training settings}
\label{subsec:training}

We performed static optimization (Figure~\ref{fig:pipelines}(i)) using the same settings as those used for PAC-NeRF.
Specifically, we trained the model for $6000$ iterations using the Adam optimizer~\cite{DPKingmaICLR2015} with learning rates of $0.1$ for the volume density and color grids and a learning rate of $0.001$ for the MLP.
The momentum terms $\beta_1$ and $\beta_2$ were set to $0.9$ and $0.999$, respectively.
In the dynamic optimization (Figure~\ref{fig:pipelines}(ii)), we trained the model for $1000$ iterations using the Adam optimizer~\cite{DPKingmaICLR2015} with a default learning rate of $6.4$ for the volume density grid.
The momentum terms $\beta_1$ and $\beta_2$ were set to $0.9$ and $0.999$, respectively.
We found that a high learning rate is useful for efficiently reducing the volume density; however, this is not necessary when the estimated mass $m$ sufficiently approaches the ground-truth mass $\hat{m}$.
Therefore, we divided the learning rate by $2$ (with a minimum of $0.1$) as long as the estimated mass $m$ was below the ground-truth mass $\hat{m}$.
Conversely, we multiplied the learning rate by $2$ (with a maximum of $6.4$) as long as the estimated mass $m$ exceeded the ground-truth mass $\hat{m}$.

We conducted volume annealing every $100$ iteration during the dynamic optimization.
When the estimated mass $m$ was significantly larger than the ground-truth mass $\hat{m}$ (specifically, when the difference exceeded $10$ in practice), the expansion process was skipped to prevent $m$ from deviating further from $\hat{m}$.

In appearance-preserving training, static optimization was performed using settings similar to those mentioned above (i.e., static optimization in Step (i) (Figure~\ref{fig:pipelines}(i))), but the number of iterations was reduced to $10$.

We empirically set the hyperparameters for the full objective $\mathcal{L}_{\text{full}}$ (Equation~\ref{eq:full}) to $\lambda_{\text{mass}} = 1$, $\lambda_{\text{pres}} = 100$, $w_{\text{depth}} = 0.01$, and $\lambda_{\text{key}} = 10$.
The hyperparameter for background loss $\mathcal{L}_{\text{bg}}$ was set to $w_{\text{bg}} = 0.2$.

\subsection{Evaluation metrics}
\label{subsec:metrics}

As mentioned in Section~\ref{subsec:problem}, we use particles $\mathcal{P}^P (t_0)$ to represent the structure (including the internal structure) of an object and estimate $\mathcal{P}^P (t_0)$ to match the ground-truth particles $\hat{\mathcal{P}}^P (t_0)$.
Therefore, we evaluated the model by measuring the distance between $\mathcal{P}^P (t_0)$ and $\hat{\mathcal{P}}^P (t_0)$ using the \textit{chamfer distance (CD)}.
The smaller the value, the higher the degree of matching.
As mentioned in Section~\ref{subsec:experiment2}, we also used the \textit{anti-chamfer distance (ACD)}, which is the chamfer distance between the predicted particles $\mathcal{P}^P(t_0)$ and ground-truth particles $\tilde{\mathcal{P}}^{P}(t_0)$, where the cavity was placed on the opposite side, to evaluate the capture of the cavity location.

\end{document}